\RequirePackage{lineno}
\documentclass[aps,pr,twocolumn,superscriptaddress,preprintnumbers,showpacs,amsmath,amssymb,floatfix]{revtex4}

\usepackage{graphicx} % Include figure files
\usepackage{dcolumn}  % Align table columns on decimal point

\usepackage{epsfig}
\usepackage{epsbox}
\usepackage[T1]{fontenc}
\usepackage{wrapfloat}
\usepackage{subfigure}
\usepackage{here}
\usepackage{array}
\usepackage{amsmath, amssymb, hhline, multirow, hangcaption, subfigure}
\usepackage{graphics}
\usepackage{graphicx}
\usepackage{lscape}
\usepackage{fancyhdr}
\usepackage{longtable}
\usepackage{supertabular}
\usepackage{refmerge}
\usepackage{dcolumn}
\usepackage{color}
\usepackage{comment}

\def \vec #1{\mbox{{\boldmath $#1$}}}

\def \B {{\cal B}}

\def \GeV {{\rm GeV}}

\def \vec #1{\mbox{{\boldmath $#1$}}}
\def \ks {K^0_S}
\begin{document}
%\linenumbers
\hspace{100mm}
\preprint{\vbox{ \hbox{   }
%                \hbox{Bellenote xxxx}
%                \hbox{Belle Preprint 2016-xx}
%                 \hbox{KEK Preprint 2016-xx}
%                \hbox{Mar. 2016}
%                 \hbox{   }
%                 \hbox{Contact: M.~Masuda, S.~Uehara, Y.~Watanabe and H.~Nakazawa}
%                 \hbox{Internal Committee:} 
%                 \hbox{\ \ S.~Eidelman (chair), T.~Matsuda, B.~Reisert }
%                 \hbox{{\bf  Version 1.0 } Oct. 28, 2012  }
%                 \hbox{Intended Journal: PRD}
%                 \hbox{\small BELLE DRAFT  VSub July 22,~2010 }
                  \hbox{Belle Preprint 2017-25}
                  \hbox{KEK Preprint 2017-36}
                  \hbox{Dec 2017}
}}
\title{ 
%\quad\\[0.5cm]  \ \\ \ \\
Study of $\ks$ pair production in single-tag
two-photon collisions\\
}

\begin{abstract}
We report a measurement of the cross section for $\ks$ 
pair production in single-tag two-photon collisions,
$\gamma^* \gamma \to \ks \ks$,
for $Q^2$ up to $30~\GeV^2$, 
where $Q^2$ is the negative of the invariant mass squared of the tagged photon. 
The measurement covers the kinematic range
$1.0~\GeV < W < 2.6~\GeV$ and $|\cos \theta^*| < 1.0$ 
for the total energy and kaon scattering angle, respectively, 
in the $\gamma^* \gamma$ center-of-mass system. 
These results are based on a data sample of 759~fb$^{-1}$ collected
with the Belle detector at the KEKB asymmetric-energy $e^+ e^-$ collider.
For the first time, the transition form factor of the $f_2'(1525)$ 
meson is measured separately for the helicity-0, -1, and -2 components 
and also compared with theoretical calculations.
Finally, the partial decay widths of the $\chi_{c0}$ and $\chi_{c2}$ mesons are 
measured as a function of $Q^2$.
\end{abstract}


\normalsize
%\author{M.~Masuda}\affiliation{Earthquake Research Institute, the University of Tokyo, Tokyo}
%\author{S.~Uehara}\affiliation{High Energy Accelerator Research Organization (KEK), Tsukuba}
%\author{H.~Nakazawa}\affiliation{National Central University, Taipei}
%\author{Y.~Watanabe}\affiliation{Kanagawa University, Kanagawa}
%\author{...}\affiliation{...}
%\collaboration{The Belle Collaboration}

%%% Paper:    g g -> KS KS
%%% Journal:  Physical Review D
%%% Contacts: M. Masuda (masudama@eri.u-tokyo.ac.jp)
%%%           S. Uehara (uehara@post.kek.jp)
%%%           Y. Watanabe (yasushi_watanabe@kanagawa-u.ac.jp)
%%%           H. Nakazawa (nkzw@post.kek.jp)
%%% Non-responding authors or those who said NO are commented out.
%%% ====================================================================
%%% Click the RELOAD button on your web browser to see the updated file.
%%% ====================================================================
%%% Use \input{author} to insert this material into your latex file.
%%%%% Force institutions to appear in alphabetical order when typeset.
\noaffiliation
%%%\affiliation{Aligarh Muslim University, Aligarh 202002}
\affiliation{University of the Basque Country UPV/EHU, 48080 Bilbao}
\affiliation{Beihang University, Beijing 100191}
%%%\affiliation{University of Bonn, 53115 Bonn}
\affiliation{Budker Institute of Nuclear Physics SB RAS, Novosibirsk 630090}
\affiliation{Faculty of Mathematics and Physics, Charles University, 121 16 Prague}
%%%\affiliation{Chiba University, Chiba 263-8522}
\affiliation{Chonnam National University, Kwangju 660-701}
\affiliation{University of Cincinnati, Cincinnati, Ohio 45221}
\affiliation{Deutsches Elektronen--Synchrotron, 22607 Hamburg}
\affiliation{University of Florida, Gainesville, Florida 32611}
%%%\affiliation{Department of Physics, Fu Jen Catholic University, Taipei 24205}
%%%\affiliation{Justus-Liebig-Universit\"at Gie\ss{}en, 35392 Gie\ss{}en}
\affiliation{Gifu University, Gifu 501-1193}
%%%\affiliation{II. Physikalisches Institut, Georg-August-Universit\"at G\"ottingen, 37073 G\"ottingen}
\affiliation{SOKENDAI (The Graduate University for Advanced Studies), Hayama 240-0193}
%%%\affiliation{Gyeongsang National University, Chinju 660-701}
\affiliation{Hanyang University, Seoul 133-791}
\affiliation{University of Hawaii, Honolulu, Hawaii 96822}
\affiliation{High Energy Accelerator Research Organization (KEK), Tsukuba 305-0801}
\affiliation{J-PARC Branch, KEK Theory Center, High Energy Accelerator Research Organization (KEK), Tsukuba 305-0801}
%%%\affiliation{Hiroshima Institute of Technology, Hiroshima 731-5193}
\affiliation{IKERBASQUE, Basque Foundation for Science, 48013 Bilbao}
%%%\affiliation{University of Illinois at Urbana-Champaign, Urbana, Illinois 61801}
\affiliation{Indian Institute of Science Education and Research Mohali, SAS Nagar, 140306}
\affiliation{Indian Institute of Technology Bhubaneswar, Satya Nagar 751007}
\affiliation{Indian Institute of Technology Guwahati, Assam 781039}
\affiliation{Indian Institute of Technology Hyderabad, Telangana 502285}
\affiliation{Indian Institute of Technology Madras, Chennai 600036}
\affiliation{Indiana University, Bloomington, Indiana 47408}
\affiliation{Institute of High Energy Physics, Chinese Academy of Sciences, Beijing 100049}
\affiliation{Institute of High Energy Physics, Vienna 1050}
\affiliation{Institute for High Energy Physics, Protvino 142281}
%%%\affiliation{Institute of Mathematical Sciences, Chennai 600113}
\affiliation{University of Mississippi, University, Mississippi 38677}
\affiliation{INFN - Sezione di Napoli, 80126 Napoli}
\affiliation{INFN - Sezione di Torino, 10125 Torino}
\affiliation{Advanced Science Research Center, Japan Atomic Energy Agency, Naka 319-1195}
\affiliation{J. Stefan Institute, 1000 Ljubljana}
\affiliation{Kanagawa University, Yokohama 221-8686}
\affiliation{Institut f\"ur Experimentelle Kernphysik, Karlsruher Institut f\"ur Technologie, 76131 Karlsruhe}
%%%\affiliation{Kavli Institute for the Physics and Mathematics of the Universe (WPI), University of Tokyo, Kashiwa 277-8583}
\affiliation{Kennesaw State University, Kennesaw, Georgia 30144}
\affiliation{King Abdulaziz City for Science and Technology, Riyadh 11442}
\affiliation{Department of Physics, Faculty of Science, King Abdulaziz University, Jeddah 21589}
\affiliation{Korea Institute of Science and Technology Information, Daejeon 305-806}
\affiliation{Korea University, Seoul 136-713}
\affiliation{Kyoto University, Kyoto 606-8502}
\affiliation{Kyungpook National University, Daegu 702-701}
\affiliation{\'Ecole Polytechnique F\'ed\'erale de Lausanne (EPFL), Lausanne 1015}
\affiliation{P.N. Lebedev Physical Institute of the Russian Academy of Sciences, Moscow 119991}
\affiliation{Faculty of Mathematics and Physics, University of Ljubljana, 1000 Ljubljana}
\affiliation{Ludwig Maximilians University, 80539 Munich}
\affiliation{Luther College, Decorah, Iowa 52101}
\affiliation{University of Malaya, 50603 Kuala Lumpur}
\affiliation{University of Maribor, 2000 Maribor}
\affiliation{Max-Planck-Institut f\"ur Physik, 80805 M\"unchen}
\affiliation{School of Physics, University of Melbourne, Victoria 3010}
%%%\affiliation{Middle East Technical University, 06531 Ankara}
\affiliation{University of Miyazaki, Miyazaki 889-2192}
\affiliation{Moscow Physical Engineering Institute, Moscow 115409}
\affiliation{Moscow Institute of Physics and Technology, Moscow Region 141700}
\affiliation{Graduate School of Science, Nagoya University, Nagoya 464-8602}
\affiliation{Kobayashi-Maskawa Institute, Nagoya University, Nagoya 464-8602}
%%%\affiliation{Nara University of Education, Nara 630-8528}
\affiliation{Nara Women's University, Nara 630-8506}
\affiliation{National Central University, Chung-li 32054}
\affiliation{National United University, Miao Li 36003}
\affiliation{Department of Physics, National Taiwan University, Taipei 10617}
\affiliation{H. Niewodniczanski Institute of Nuclear Physics, Krakow 31-342}
\affiliation{Nippon Dental University, Niigata 951-8580}
\affiliation{Niigata University, Niigata 950-2181}
%%%\affiliation{University of Nova Gorica, 5000 Nova Gorica}
\affiliation{Novosibirsk State University, Novosibirsk 630090}
\affiliation{Osaka City University, Osaka 558-8585}
%%%\affiliation{Osaka University, Osaka 565-0871}
\affiliation{Pacific Northwest National Laboratory, Richland, Washington 99352}
\affiliation{Panjab University, Chandigarh 160014}
%%%\affiliation{Peking University, Beijing 100871}
\affiliation{University of Pittsburgh, Pittsburgh, Pennsylvania 15260}
%%%\affiliation{Punjab Agricultural University, Ludhiana 141004}
%%%\affiliation{Research Center for Electron Photon Science, Tohoku University, Sendai 980-8578}
%%%\affiliation{Research Center for Nuclear Physics, Osaka University, Osaka 567-0047}
\affiliation{Theoretical Research Division, Nishina Center, RIKEN, Saitama 351-0198}
\affiliation{RIKEN BNL Research Center, Upton, New York 11973}
%%%\affiliation{Saga University, Saga 840-8502}
\affiliation{University of Science and Technology of China, Hefei 230026}
%%%\affiliation{Seoul National University, Seoul 151-742}
%%%\affiliation{Shinshu University, Nagano 390-8621}
\affiliation{Showa Pharmaceutical University, Tokyo 194-8543}
\affiliation{Soongsil University, Seoul 156-743}
\affiliation{University of South Carolina, Columbia, South Carolina 29208}
\affiliation{Stefan Meyer Institute for Subatomic Physics, Vienna 1090}
\affiliation{Sungkyunkwan University, Suwon 440-746}
%%%\affiliation{School of Physics, University of Sydney, New South Wales 2006}
\affiliation{Department of Physics, Faculty of Science, University of Tabuk, Tabuk 71451}
\affiliation{Tata Institute of Fundamental Research, Mumbai 400005}
%%%\affiliation{Excellence Cluster Universe, Technische Universit\"at M\"unchen, 85748 Garching}
\affiliation{Department of Physics, Technische Universit\"at M\"unchen, 85748 Garching}
\affiliation{Toho University, Funabashi 274-8510}
%%%\affiliation{Tohoku Gakuin University, Tagajo 985-8537}
\affiliation{Department of Physics, Tohoku University, Sendai 980-8578}
\affiliation{Earthquake Research Institute, University of Tokyo, Tokyo 113-0032}
\affiliation{Department of Physics, University of Tokyo, Tokyo 113-0033}
\affiliation{Tokyo Institute of Technology, Tokyo 152-8550}
\affiliation{Tokyo Metropolitan University, Tokyo 192-0397}
%%%\affiliation{Tokyo University of Agriculture and Technology, Tokyo 184-8588}
\affiliation{University of Torino, 10124 Torino}
%%%\affiliation{Utkal University, Bhubaneswar 751004}
\affiliation{Virginia Polytechnic Institute and State University, Blacksburg, Virginia 24061}
\affiliation{Wayne State University, Detroit, Michigan 48202}
\affiliation{Yamagata University, Yamagata 990-8560}
\affiliation{Yonsei University, Seoul 120-749}
  \author{M.~Masuda}\affiliation{Earthquake Research Institute, University of Tokyo, Tokyo 113-0032} % NPC
  \author{S.~Uehara}\affiliation{High Energy Accelerator Research Organization (KEK), Tsukuba 305-0801}\affiliation{SOKENDAI (The Graduate University for Advanced Studies), Hayama 240-0193} % KEK
  \author{Y.~Watanabe}\affiliation{Kanagawa University, Yokohama 221-8686} % Kanagawa
% \author{A.~Abdesselam}\affiliation{Department of Physics, Faculty of Science, University of Tabuk, Tabuk 71451} % Tabuk
  \author{I.~Adachi}\affiliation{High Energy Accelerator Research Organization (KEK), Tsukuba 305-0801}\affiliation{SOKENDAI (The Graduate University for Advanced Studies), Hayama 240-0193} % KEK
% \author{K.~Adamczyk}\affiliation{H. Niewodniczanski Institute of Nuclear Physics, Krakow 31-342} % Krakow
  \author{J.~K.~Ahn}\affiliation{Korea University, Seoul 136-713} % Korea
  \author{H.~Aihara}\affiliation{Department of Physics, University of Tokyo, Tokyo 113-0033} % Tokyo
  \author{S.~Al~Said}\affiliation{Department of Physics, Faculty of Science, University of Tabuk, Tabuk 71451}\affiliation{Department of Physics, Faculty of Science, King Abdulaziz University, Jeddah 21589} % Tabuk
% \author{K.~Arinstein}\affiliation{Budker Institute of Nuclear Physics SB RAS, Novosibirsk 630090}\affiliation{Novosibirsk State University, Novosibirsk 630090} % BINP
% \author{Y.~Arita}\affiliation{Graduate School of Science, Nagoya University, Nagoya 464-8602} % Nagoya
  \author{D.~M.~Asner}\affiliation{Pacific Northwest National Laboratory, Richland, Washington 99352} % PNNL
  \author{H.~Atmacan}\affiliation{University of South Carolina, Columbia, South Carolina 29208} % SouthCarolina
  \author{V.~Aulchenko}\affiliation{Budker Institute of Nuclear Physics SB RAS, Novosibirsk 630090}\affiliation{Novosibirsk State University, Novosibirsk 630090} % BINP
% ...
% Ljubljana
% \author{N.~Zwahlen}\affiliation{\'Ecole Polytechnique F\'ed\'erale de Lausanne (EPFL), Lausanne 1015} % Lausanne
\collaboration{The Belle Collaboration}

%\vspace*{\baselineskip}
\medskip
%\date{November 21, 2017, Ver.~4.2}

\noaffiliation

\maketitle
%\newpage

%%%% >>>> keep the final version single-spaced
\tighten

%{\renewcommand{\thefootnote}{\fnsymbol{footnote}}}
%\setcounter{footnote}{0}

%\begin{multicols}{2}

\section{Introduction}
\label{sec:intro}
Single-tag two-photon production of a hadron pair, 
$\gamma^* \gamma \to h h'$,
provides valuable information on the nature of hadrons 
by exploiting an additional degree of freedom, $Q^2$, 
which is the negative of the invariant mass squared of the tagged photon. 
These processes can be studied through the reaction 
$e^+ e^- \to e^\pm (e^\mp) h h'$, 
where $(e^\mp)$ implies an undetected electron or positron, 
and provide vital input on 
hadron structure and properties, in the context of Quantum Chromodynamics (QCD). 

In the framework of perturbative QCD, Kawamura and Kumano, using
generalized quark distribution amplitudes, emphasized
the importance of exclusive production 
in single-tag two-photon processes as a way to unambiguously identify the
nature of exotic hadrons~\cite{kawamura}.
They showed, for example, that studies of  
$\gamma^* \gamma \to h\bar{h}$, where $h$ is the $f_0(980)$ or the $a_0(980)$ meson, could clearly reveal whether
the $f_0(980)$ and the $a_0(980)$ states were tetraquarks. In addition, 
a data-driven dispersive approach was suggested that allows a more 
precise estimate of the hadronic light-by-light contribution to the 
anomalous magnetic moment of the muon ($g-2$)~\cite{Colangelo, Colangelo2}.

Recently, we have performed a measurement of the differential cross section for
single-tag two-photon production of $\pi^0 \pi^0$~\cite{masuda}.
There, we derived for the first time the transition form factor (TFF) of both
the $f_0(980)$ and the $f_2(1270)$ mesons for helicity-0, -1, and -2 components
at $Q^2$ up to 30 GeV$^2$.

In this paper, we report a measurement of the process
 $e^+ e^- \to e^\pm (e^\mp) \ks\ks$, 
where one of the $e^\pm$ is detected together with $\ks \ks$,
while the other $e^\mp$ is scattered in the forward direction
and undetected.

\begin{figure}
 \centering
{\epsfig{file=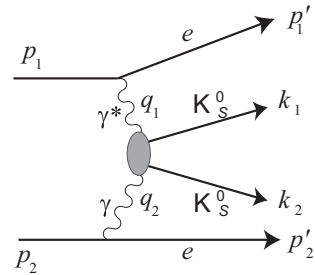,width=40mm}}
 \caption{Feynman diagram for the process 
$e^+ e^- \to e (e) \ks \ks$
and definition of the eight four-momenta.}
\label{fig:eeksks}
\end{figure}

\begin{figure}
 \centering
  {\epsfig{file=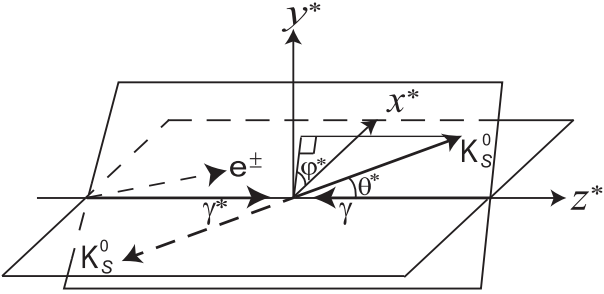,width=80mm}}
 \caption{
Definition of the $\gamma^* \gamma$ c.m. coordinate system for $\gamma^* \gamma \to \ks \ks$. 
The ``incident'' $\gamma^*$ has momentum along
the $z^*$ axis with $p_{z^*}>0$, the tagging $e^{\pm}$ is in the
$x^*z^*$ plane with $p_{{\rm tag}\, x^*}>0$, and the forward-going
$\ks$ (\textit{i.e.,} having $p_{Kz^*} > 0$) is produced at angles ($\theta^*$,  $\varphi^*$).
}
\label{fig:coord}
\end{figure}

A Feynman diagram for the process of interest is shown in Fig.~\ref{fig:eeksks},
where the four-momenta of particles involved are defined.
We consider the process  $\gamma^* \gamma \to \ks \ks$
in the center-of-mass (c.m.) system of the $\gamma^* \gamma$.
We define the $x^*y^*z^*$-coordinate system as shown in 
Fig.~\ref{fig:coord} at
fixed $W$ and $Q^2$, where $W$ is the total energy in the $\gamma^* \gamma$ c.m. frame. 
One of the $\ks$ mesons is scattered at polar angle $\theta^*$ and azimuthal angle $\varphi^*$.
Since the final-state particles are identical, only the region where 
$\theta^* \le \pi/2$ and $0 \le |\varphi^*| \le \pi$ is of interest.
The $z^* $axis is defined along the incident $\gamma^*$ and the $x^* z^*$ plane 
is defined by the detected tagging $e^\pm$ such that 
$p_{{\rm tag}\, x^*} > 0$, where $\vec{p}_{\rm tag}$ is the three-momentum
of the tagging $e^\pm$.

The differential cross section for
$\gamma^* \gamma \to \ks \ks$ 
taking place at an $e^+e^-$ collider 
is calculated using the
helicity-amplitude formalism as follows~\cite{masuda,gss}: 

\begin{equation}
\frac{d \sigma(\gamma^* \gamma \to \ks \ks)}{d \Omega} = 
\sum_{n=0}^2 t_n \cos (n  \varphi^*),
\label{eqn:dsdo}
\end{equation}
with
\begin{eqnarray}
t_0 &=& 
|M_{++}|^2 + |M_{+-}|^2
+ 2 \epsilon_0 |M_{0+}|^2 , \label{eqn:t0}\\
t_1 &=& 2 \epsilon_1 \Re \left[ (M_{+-}^* - M_{++}^*) M_{0+} \right],
\label{eqn:t1}\\
t_2 &=& - 2 \epsilon_0 \Re (M_{+-}^* M_{++}) ,
\label{eqn:t2}
\end{eqnarray}
where $M_{++}$, $M_{0+}$, and $M_{+-}$ are separate helicity amplitudes;
$+, -, 0$ indicate the helicity state of the incident virtual photon
along, opposite, or transverse to the quantization axis, respectively,
and $\epsilon_0$ and $\epsilon_1$ are given by
\begin{eqnarray}
\epsilon_0 &=& \frac{1-x}{1 -  x + \frac{1}{2}x^2 } , 
\label{eqn:eps0}\\
\epsilon_1 &=& \frac{(2 - x)\sqrt{\frac{1}{2}(1-x)}}{1 -  x + \frac{1}{2}x^2 } .
\label{eqn:eps1}
\end{eqnarray}
Here, $x$ is defined as
\begin{equation}
x = \frac{q_1 \cdot q_2}{p_1 \cdot q_2} ,
\label{eqn:defx}
\end{equation}
where $q_1, q_2$, and $p_1$ are the four-momenta of the virtual and real photons 
and an incident lepton, respectively, as defined in 
Fig.~\ref{fig:eeksks}.
When Eq.~(\ref{eqn:dsdo}) is integrated over $\varphi^*$, we obtain
\begin{equation}
\frac{d \sigma(\gamma^* \gamma \to \ks \ks)}{4\pi d |\cos \theta^*|} 
= |M_{++}|^2 + |M_{+-}|^2 + 2 \epsilon_0 |M_{0+}|^2 .
\label{eqn:dsdcos}
\end{equation}
The total cross section is obtained by integrating Eq.~(\ref{eqn:dsdcos}) 
over $\cos \theta^*$, and
can be written as
\begin{equation}
\sigma_{\rm tot}(\gamma^* \gamma \to \ks \ks)
= \sigma_{TT} +  \epsilon_0 \sigma_{LT},
\label{eqn:tots}
\end{equation}
where $\sigma_{TT}$ ($\sigma_{LT}$) corresponds to the total cross
section in which both photons are transversely polarized 
(one photon is longitudinally polarized and the other is transversely polarized).

A $\ks$ pair produced in the final state of the process
 $e^+ e^- \to e^\pm (e^\mp) \ks\ks$
is a pure $C$-even state and has no contribution
from single-photon production (``bremsstrahlung process''),
whose effect must otherwise be considered in two-photon production 
of $K^+ K^-$.

Schuler, Berends, and van Gulik (SBG) have calculated mesonic TFFs based 
on the heavy-quark approximation~\cite{schuler}.
They found that their calculations were also applicable to light
mesons with only minor modifications.
The predicted $Q^2$ dependence of the TFFs
for mesons with $J^{PC} = 0^{++}$ and $2^{++}$
is summarized in Table~\ref{tab:pred}, where $W$ is 
replaced by the equivalent mass $M$.
\begin{center}
\begin{table}[htb]
\caption{Predicted $Q^2$ dependence of mesonic transition form factor
for various helicities of the two colliding photons~\cite{schuler}.
Each term has a common factor of
$(1 + Q^2/M^2)^{-2}$. 
}
\label{tab:pred}
\begin{tabular}{c|ccc} \hline \hline
&&&\\[-10pt]
$J^{PC}$ & \multicolumn{3}{c}{$Q^2$ dependence 
$\left[ \div \left( 1 + \frac{Q^2}{M^2} \right)^2 \right]$ } 
\\ \cline{2-4}
&&&\\[-10pt]
& helicity-0 & helicity-1 & helicity-2 \\ \hline\hline
&&&\\[-10pt]
$0^{++}$ & $\left( 1 + \frac{Q^2}{3 M^2} \right)$
& -- & -- \\ 
&&&\\[-10pt]
\hline
&&&\\[-10pt]
$2^{++}$ & $\frac{Q^2}{\sqrt{6} M^2}$ & 
$\frac{\sqrt{Q^2}}{\sqrt{2}M}$ & 1 \\
&&&\\[-10pt]
 \hline \hline
\end{tabular}
\end{table}
\end{center}
\nobreak

In this paper, we report a measurement of 
$\gamma^* \gamma \to \ks \ks$, extracting for the first time 
the $Q^2$ dependence of the production cross section in the charmonium
mass region (specifically for the $\chi_{c0}$
and $\chi_{c2}$ mesons), near the $\ks \ks$ mass threshold, and also
the separate helicity-0, -1, and -2 TFF of the $f_2'(1525)$ meson up 
to $Q^2 = 30~\GeV^2$. 
These measurements complement our earlier measurements for
the corresponding no-tag process $\gamma \gamma \to \ks \ks$ over the range 
$1.05~\GeV \le W \le 4.0~\GeV$~\cite{ksks}.

\section{Experimental apparatus and Data Sample}
\label{sec:exper}
We use a 759~fb$^{-1}$ data sample recorded with
the Belle detector~\cite{belle, belleptep} at the KEKB asymmetric-energy
$e^+e^-$ collider~\cite{kekb, kekb2}; this data sample is identical to that used for the
previous $\gamma^*\gamma\to\pi^0 \pi^0$ measurement~\cite{masuda}.

\subsection{Belle detector}
\label{sub:belle}
A comprehensive description of the Belle detector is
given elsewhere~\cite{belle, belleptep}. 
In the following, we describe only the
detector components essential for this measurement.
Charged tracks are reconstructed from the drift-time information in a central
drift chamber (CDC) located in a uniform 1.5~T solenoidal magnetic field.
The $z$ axis of the detector and the solenoid is opposite the positron beam.
The CDC measures the longitudinal and transverse momentum components, 
{\it i.e.}, along the $z$ axis and in the $r\varphi$ plane perpendicular to
the beam, respectively.
The trajectory coordinates near the collision point are measured by a
silicon vertex detector.  
A barrel-like arrangement of time-of-flight (TOF) counters 
is used to supplement the CDC trigger for charged particles and to measure
their time of flight.  
Charged-particle identification (ID) is achieved by including information from 
the CDC, the TOF, and an array of aerogel threshold Cherenkov counters.  
Photon detection and energy measurements are performed with a CsI(Tl) 
electromagnetic calorimeter (ECL)
by clustering the ECL energy deposits not matched to extrapolated CDC charged
track trajectories.
Electron identification is based on $E/p$, the ratio of the ECL
calorimeter energy to the CDC track momentum.

\subsection{Triggers}
\label{sub:sample}
The triggers that are important for this analysis are
the ECL-based~\cite{ecltrig} HiE (High-energy threshold) trigger and the Clst4 (four-energy-cluster) 
energy triggers.
The HiE trigger requires that the sum of the energies measured 
by the ECL in an event exceed 1.15~GeV, 
but that the event topology not be similar to Bhabha scattering (``Bhabha veto''); 
the latter requirement is enforced by the absence of the CsiBB trigger,
which is designed to identify back-to-back Bhabha events~\cite{ecltrig}.
The Clst4 trigger requires at least four separated energy clusters
in the ECL with each cluster energy above 0.11~GeV;
this trigger is not vetoed by the CsiBB.
Five clusters are expected in total in the 
signal events of interest if all the final-state particles 
are detected within the fiducial volume of the ECL trigger ($18.5^{\circ} < \theta < 128.6^{\circ}$).

Belle employs many distinct track triggers that
require anywhere from two to four CDC tracks, in conjunction with
pre-specified TOF and/or ECL information.
Among these track triggers, the Bhabha veto is applied 
to the two-track triggers only.

The candidate signal topology nominally has five tracks
and one high-energy cluster from the electron. 
Over the entire kinematic range of interest, 
the trigger efficiency is in general quite high,
owing to the trigger requirements demanding two or three CDC 
tracks with TOF and ECL hits, with the exception of the
lowest $Q^2$ region probed in this analysis, 
where the particles tend to scatter into very small polar-angle regions. 
The typical trigger efficiency is 95\%, with slightly lower efficiency 
(around 90\%) for events having both $W\le 1.5~\GeV$ and $Q^2\le 5~\GeV^2$.

\subsection{Signal Monte Carlo}
We use the signal Monte Carlo (MC) generator, TREPSBSS, which has been
developed to calculate the efficiency for single-tag two-photon events,
$e^+e^- \to e(e)X$, as well as the two-photon luminosity function for
$\gamma^* \gamma$ collisions at an $e^+e^-$ collider 
, following our previous $\pi^0\pi^0$ study~\cite{masuda,treps}.

We choose fifteen different $W$ points between 1.0~GeV and 3.556~GeV,
including two $\chi_{cJ}$ ($J$ =0, 2) mass points,
for the calculation of the luminosity function and event generation.
The luminosity function is defined as the conversion factor
from the $e^+e^-$-based
differential cross section, $d^2 \sigma_{ee}/ dW dQ^2$, to the
$\gamma^*\gamma$-based cross section, $\sigma(W, Q^2)$~\cite{masuda}.
The scattering angle of the $\ks$ 
is uniformly distributed in the
$\gamma^* \gamma$ c.m. system in the MC sample.
To properly weight our MC sample by the beam-energy distributions used
for the data analysis, we generate $4 \times 10^5$ events [$8 \times 10^4$ events] for
the beam energy point of $\Upsilon(4S)$ [$\Upsilon(5S)$].

We use a GEANT3-based detector simulation~\cite{geant}
to study the propagation of the generated particles and their daughters through the detector.
The $\ks$ pairs decay generically in the detector simulator.
The same code used for analysis of true 
data is used for reconstruction and selection of the MC simulated events.

\section{Event Selection}
\label{sec:evsel}
Event selection parallels that of our previous
$\pi^0 \pi^0$ analysis~\cite{masuda}.
Here, we also present comparisons between data and simulation for our selected 
$\ks \ks$ samples.

\subsection{Selection criteria}
\label{sub:selec}
A candidate $e^+ e^- \to e(e) \ks \ks$ signal event 
with $\ks$ decaying to $\pi^+ \pi^-$  
contains an energetic tagging electron and four charged pions.
The kinematic variables are calculated in the laboratory system 
unless otherwise noted; those in the $e^+e^-$ or $\gamma^* \gamma$
c.m. frame are identified with an asterisk in this section.

We require exactly five tracks satisfying $p_t > 0.1~\GeV/c$,
$dr < 5$~cm, and $|dz| < 5$~cm. 
Among these, at least two tracks must satisfy $p_t > 0.4~\GeV/c$,
$-0.8660 < \cos \theta <0.9563$, and $dr < 1$~cm.
Here, $p_t$ is the transverse momentum in the laboratory frame,
$\theta$ is the polar angle of the momentum, 
and ($dr$, $dz$) are the cylindrical coordinates of the point of closest 
approach of the track to the nominal $e^+e^-$ primary interaction
point; all four variables are measured with respect to 
the $z$ axis.

One of the tracks having $p_t > 0.4$~GeV/$c$ and $p > 1.0$~GeV/$c$ 
must also be electron- (or positron-) like. 
This is ensured by
requiring that the ratio of the candidate calorimeter cluster energy,
using the cluster-energy correction outlined previously~\cite{masuda},
relative to the absolute momentum satisfy $E/p>0.8$.

We search for exactly two $\ks$ candidates, each of which is 
reconstructed from a unique charged-pion pair.
Each pion satisfies the $K/\pi$ particle ID separation criterion 
${\cal L}_K/({\cal L}_K+{\cal L}_\pi) < 0.8$, 
which is applied for the likelihood probability ratio for the hadron identification hypotheses 
obtained by combining information 
from the particle-ID detectors. 
The invariant mass of the $\ks$ candidates 
at the reconstructed decay vertex must be within $\pm 20$~MeV/$c^2$ of the 
nominal $\ks$ mass, 0.4976~GeV/$c^2$ \cite{pdg2016}. 

After the two $\ks$ candidates are found, we refine the event
selection by additionally requiring that
the average of, and difference between,
the masses of the two $\ks$'s be 
within $\pm 5$~MeV/$c^2$ from the nominal $\ks$ mass, 
and smaller than 10~MeV/$c^2$, respectively~\cite{ksks}. 
Each $\ks$ decay vertex must lie within the cylindrical volume defined by
$0.3\,{\rm cm} < r_{VK} < 8\,{\rm cm}$ 
and  $-5\,{\rm cm} < z_{VK} < +7\,{\rm cm}$, 
where ($r_{VK}$, $z_{VK}$) is the decay-vertex position of the $\ks$.

We do not require the characteristic relation between 
the $z$ component of the observed total momenta and the charge of 
the tagging lepton that was used in the previous $\pi^0\pi^0$ analysis~\cite{masuda}, as
this results in no effective additional background reduction;
the background from $e^+e^-$ annihilation
is already very small, given our distinctive event topology.

We apply an acoplanarity cut between the c.m. momenta of the electron
and the two-$\ks$ system, namely, that
their opening angle projected onto 
the $r \varphi$ plane must exceed $\pi - 0.1$ radians.

Finally, we apply kinematic selection using the $E_{\rm ratio}$ and
$p_t$-balance variables just as was done for the $\pi^0 \pi^0$ selection~\cite{masuda}.
Those definitions of $E_{\rm ratio}$ and $p_t$ balance  
are reproduced here for completeness.
The energy ratio is 
\begin{equation}
E_{\rm ratio} = \frac{E^{* {\rm measured}}_{\ks\ks}}
{E^{* {\rm expected}}_{\ks\ks}},
\label{eqn:eratio}
\end{equation}
where $E^{* {\rm measured}}_{\ks\ks}$ ($E^{* {\rm expected}}_{\ks\ks}$) is 
the $e^+ e^-$ c.m. energy of the $\ks \ks$ system measured directly 
(as expected by kinematics, assuming no radiation). 
The $p_t$-balance $|\Sigma \vec{p}_t^*|$ is defined by 
\begin{equation}
|\Sigma \vec{p}_t^*| =
|\vec{p}^*_{t,e} + \vec{p}^*_{t,K1} + \vec{p}^*_{t, K2} | .
\end{equation}
We require that the quadratic combination of the two variables
($E_{\rm ratio}$ and $|\Sigma \vec{p}_t^*|$) satisfy
\begin{equation}
\sqrt{ \left( \frac{E_{\rm ratio}-1}{0.04} \right) ^2 +
\left( \frac{|\Sigma \vec{p}_t^*|}{0.1~\GeV /c} \right) ^2 } \le 1.
\end{equation}

\label{sub:assign}
We assign four kinematic variables --- $Q^2$, $W$, $|\cos \theta^*|$, and
$|\varphi^*|$ --- to each candidate event, similar to  
the $\pi^0 \pi^0$ analysis~\cite{masuda}.

\subsection{Distributions of the signal candidates and comparison with 
the signal-MC events}
\label{sub:datamc}
In this subsection, we present various distributions of the
selected signal candidates. 
The backgrounds are expected to be quite low in the 
experimental data. 
Some of the data distributions are compared with those of the signal-MC 
samples, where a uniform angular 
distribution and a representative $Q^2$ dependence~\cite{masuda} are assumed.

\begin{figure}
\centering
\includegraphics[width=8.0cm]{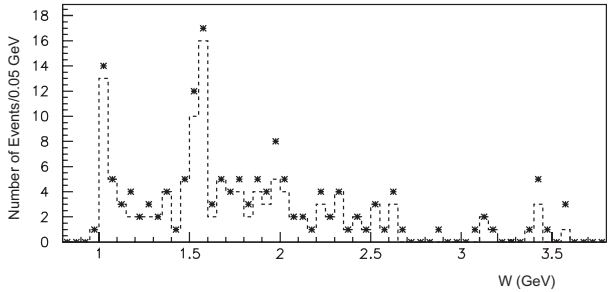}
\centering
\caption{The experimental $W$ distributions of the
signal candidates at 2~GeV$^2$ (3~GeV$^2$) $< Q^2 < 30$~GeV$^2$ 
as indicated by the asterisks (dashed histogram).
Backgrounds have not been subtracted. 
}
\label{fig:wdist}
\end{figure}

 The experimental $W$ distribution for events passing our
selection criteria is shown in Fig.~\ref{fig:wdist} for $W \le$ 3.8~GeV.
A structure corresponding to the tensor 
$f'_2(1525)$ resonance is clearly visible. We also note
an apparent enhancement
near the 
$K^0_SK^0_S$ mass threshold, that may be associated 
with the $f_0(980)$ and/or the $a_0(980)$ mesons.
 We find 124 (14) events in the region $W < 3.0~\GeV$ and
$3~\GeV^2 < Q^2 < 30~\GeV^2$ ($3.0~\GeV < W < 3.8~$GeV and
$2~\GeV^2 < Q^2 < 30~\GeV^2$).

We now focus on events having $W \le 2.6$~GeV
and the two $\chi_{cJ}$($J=0, 2$) mass regions, where we detect
the signal process with a high efficiency and a good signal-to-noise ratio. 
For the same reason, we also constrain the $Q^2$ region to 
3~GeV$^2 \le Q^2 \le 30$~GeV$^2$ (2~GeV$^2 \le Q^2 \le 30$~GeV$^2$) 
for $W \le 2.6$~GeV (the $\chi_{cJ}$ mesons).  

\begin{figure}
\centering
\includegraphics[width=8cm]{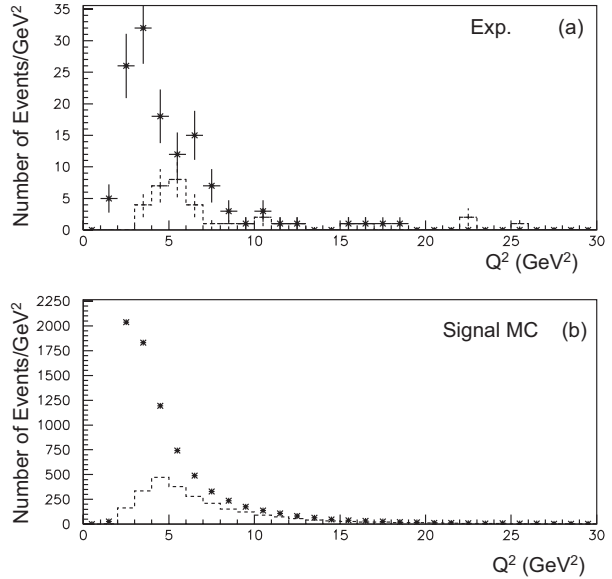}
\centering
\caption{(a) The $Q^2$ distributions for
the data samples at $W \le 4.0$~GeV.
The asterisks and the dashed histogram 
are for the p-tag and e-tag samples, respectively.
(b) The corresponding 
distributions from the signal MC events.
Statistics of the MC figures are arbitrary, 
but the scale is common for the e- and p-tags in each panel,
so that their ratio can be compared between MC and data.
}
\label{fig:q2dist}
\end{figure}

\begin{figure*}
\centering
\includegraphics[width=14cm]{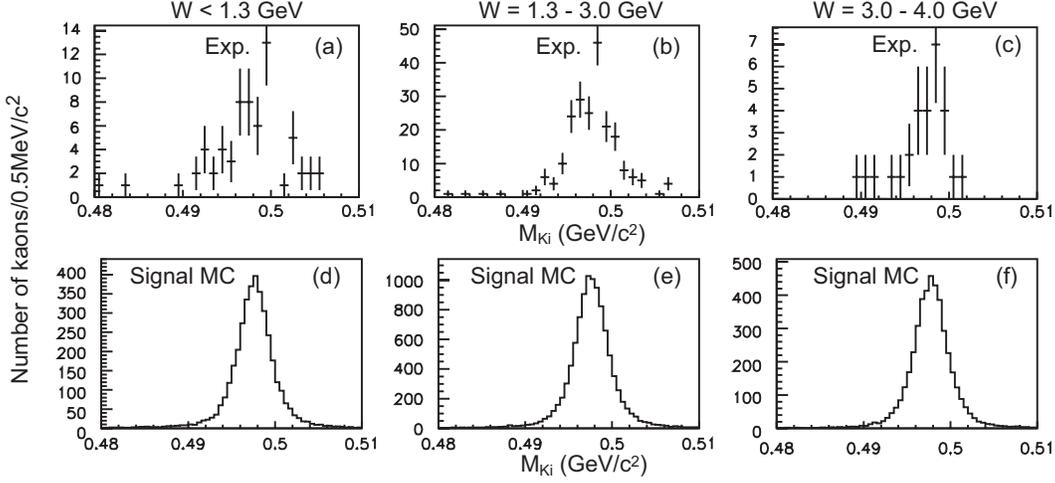}
\centering
\caption{(a,b,c) 
Reconstructed $\pi^+ \pi^-$ invariant mass, as measured
at each determined decay vertex for
the data, in three different $W$ ranges, as indicated
above each panel.
(d,e,f) The corresponding distributions from the signal MC.
}
\label{fig:ksmass}
\end{figure*}

\begin{figure*}
\centering
\includegraphics[width=14cm]{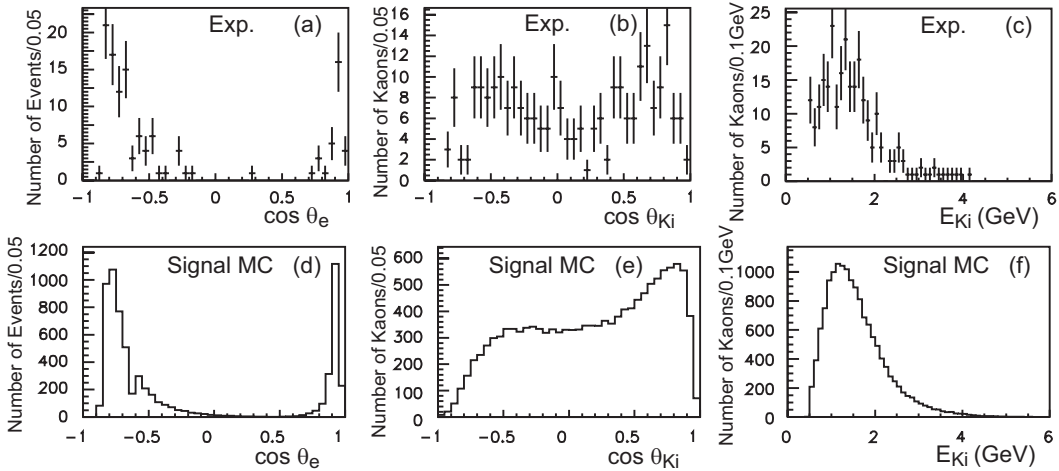}
\centering
\caption{
The distributions for experimental signal candidates (top row) and signal MC (bottom row) 
for (a,d) the cosine of the laboratory polar
angle of the tagging electron, (b,e) the cosine of the laboratory polar angle of the two $\ks$ candidates 
(two entries per event), and (c,f) the laboratory energy of the two $\ks$ candidates (two entries
per event).
}
\label{fig:ekang_ene}
\end{figure*}

For comparison, the corresponding distributions from the signal MC in this kinematic regime
are shown in Figs.~\ref{fig:q2dist} -- \ref{fig:ekang_ene}.
In our analysis, we sometimes differentiate electron-tag (e-tag) from positron-tag (p-tag)
to facilitate studies of systematics.
We find that the p-tag has a much higher efficiency than that of the e-tag
in the lowest $Q^2$ region, where the cross section is large 
(Fig.~\ref{fig:q2dist}).

Figure~\ref{fig:ksmass} compares the measured distributions of the 
reconstructed  $\pi^+ \pi^-$ invariant mass at each $\ks$-candidate decay 
vertex with MC in three different $W$ ranges, as indicated above each panel pair. 
All the selection criteria, except those related to the reconstructed
$\ks$ invariant masses ($M_{Ki}$), have been applied to the sample. 
Non-$\ks$ background is seen to be small.

Figure~\ref{fig:ekang_ene} shows the cosine of the polar 
angle of the tagging electron, that of the neutral kaon, and the energy 
of the neutral kaon in the laboratory frame for the sample at $W<3.0$~GeV 
and 3~GeV$^2 \le Q^2 \le 30$~GeV$^2$. 
They all show satisfactory agreement, given the approximate
$Q^2$ and isotropic angular dependence in the signal-MC sample.

\begin{figure}
\centering
\includegraphics[width=8cm]{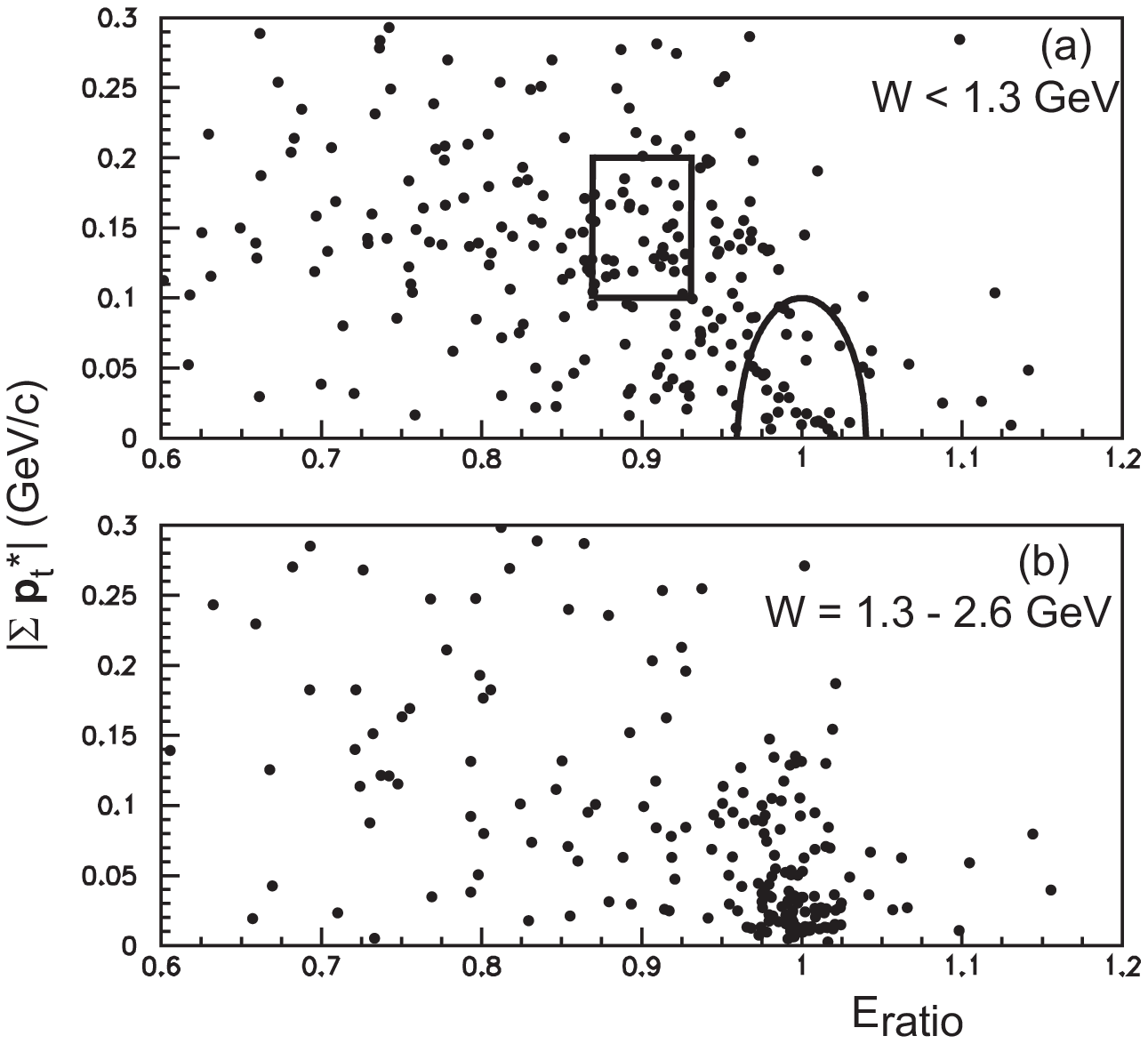}
\centering
\caption{Distribution of $p_t$ balance
vs. $E_{\rm ratio}$ for the experimental
samples to which the selection criteria
other than those related to the illustrated variables have
been applied. The $W$ region for the samples
is shown in each panel.  The half-ellipse
and the rectangle in (a) show the signal and
control regions, respectively.
}
\label{fig:nonex_up}
\end{figure}

Two-dimensional plots for $p_t$ balance ($|\Sigma \vec{p}_t^*|$) vs. $E_{\rm ratio}$ 
are shown in Fig.~\ref{fig:nonex_up}.
We find that there are backgrounds with a slightly smaller $E_{\rm ratio}$
and slightly larger $p_t$ imbalance for the data at $W < 1.3$~GeV.
These are considered to arise from the non-exclusive backgrounds
$\gamma^* \gamma \to \ks \ks X$, where $X$ is a $\pi^0$ or some combination
of otherwise undetected particles.
We discuss and subtract the background contamination of
this component in the next sections.
Such a large background contamination is not observed for $W > 1.3$~GeV.

\section{Background estimation}
\label{sec:bkgest}

\subsection{Non-\boldmath{$\ks$} background processes}
Backgrounds may arise from events in which there are either zero or only 
one true $\ks$. The latter may include contributions from 
$\ks K^\pm \pi^\mp$.
The backgrounds from these processes 
are expected to be largely eliminated by 
requirements on the invariant mass and flight length for 
each of the neutral kaon candidates.

\begin{figure}
\centering
\includegraphics[width=7cm]{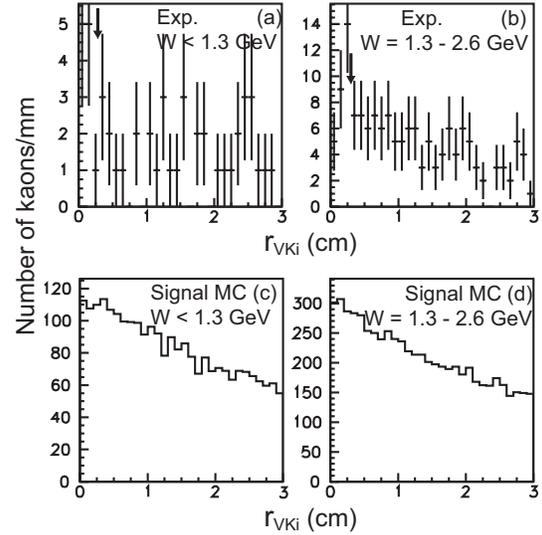}
\centering
\caption{(a,b) Experimental distribution of $r_{VKi}$ ($r$ coordinate of the $\pi^+\pi^-$ vertex 
point for a $\ks$ candidate) for an event in which the other 
kaon-vertex coordinate satisfies the selection criterion $r_{VKj} > 0.3$~cm. 
The $W$ region for each sample is shown in each panel. 
The vertical arrows indicate the selection criterion.
(c,d) The corresponding distributions from the signal-MC samples. 
Statistics of the MC figures are arbitrary.
}
\label{fig:rvki_r}
\end{figure}

If such a background component were present in the data, we would expect
an event concentration
at $r_{VKi}<0.2$~cm, based on studies of  
non-resonant $\pi^+\pi^-\pi^+\pi^-$ and $\ks K^\pm \pi^\mp$ processes, using
both the MC and background-enriched data samples.
In Figs.~\ref{fig:rvki_r}(a) and \ref{fig:rvki_r}(b), 
we show the distribution of $r_{VKi}$ for the case
$r_{VKj}>0.3$~cm ($j \neq i$)
for experimental events where the criteria other than $r_{VKi}$ have been 
applied, separately for the two $W$ regions.
These are consistent with the signal-MC distributions
shown in Figs.~\ref{fig:rvki_r}(c) and \ref{fig:rvki_r}(d). 
According to this study,
the background from this source is estimated to be
less than one event in the entire data sample, 
so we neglect its contribution.

\subsection{Non-exclusive background processes}
\label{sub:other}
The non-exclusive background processes,
$e^+e^- \to e (e) \ks \ks X$, where $X$ denotes one or multiple hadrons, 
are in general subdivided into two-photon ($C$-even) and
virtual pseudo-Compton (bremsstrahlung, $C$-odd) processes,
although these may interfere with each other if the same $X$
is allowed for both processes. 
The majority of such background events populate the
small-$E_{\rm ratio}$ and large-$p_t$ imbalance region, 
{\it e.g.}, $(E_{\rm ratio} < 0.8)~\cap~(|\Sigma \vec{p}_t^*| > 0.1$~GeV/$c$).
This feature is distinct from the aforementioned background processes 
that can populate the region near $E_{\rm ratio} = 1$ 
and peak near $|\Sigma \vec{p}_t^*| = 0$.

\begin{figure}
\centering
\includegraphics[width=7cm]{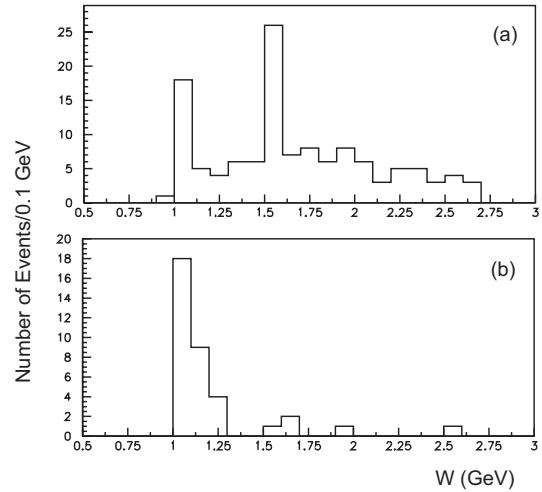}
\centering
\caption{
The $W$ distribution of experimental-data events in 
(a) the signal region and (b) the control region.
}
\label{fig:nonex_lo}
\end{figure}

To further assess background contributions,
we consider the correlation between these two variables in
the experimental sample, as illustrated in Fig.~\ref{fig:nonex_up}.
We estimate the relative ratio of the number of non-exclusive background
events to the signal yield by counting the number of events in the control 
region outside the signal region, that is, 
$(0.87 < E_{\rm ratio} < 0.93)
\cap  (0.1~\GeV/c < |\Sigma \vec{p}_t^*| < 0.2~\GeV/c)$
where the background component would be relatively large,
as well as in the selected signal region [Fig.~\ref{fig:nonex_up}(a)].
The $W$ dependence of the number of events thus obtained
in the signal and control regions is shown in Figs.~\ref{fig:nonex_lo}(a) and \ref{fig:nonex_lo}(b),
respectively. 
The peak in the 1.0 -- 1.2~GeV region for the control samples 
implies that the signal samples include a significant background
in the same $W$ region. 

We generate background $\ks \ks \pi^0$ final-state MC events,
which distribute uniformly in phase space, to estimate the 
background contamination in the signal region. 
The estimation using this process, which corresponds to the minimum
particle multiplicity of $X$, leads to a conservative (\textit{i.e.,} on the larger side)
estimate for the background fraction, since such backgrounds tend
to distribute themselves close to the signal region.

 The expected ratio of the background magnitude in the
signal region to that in the control region ($f_{\rm bs}$) 
is 13\%. We also estimate the ratio of the signal events falling 
in the control region to
that in the signal region ($f_{\rm sc}$) to be 5.6\%. 
We determine the expected background-component ratio
in the sample in the signal region, $f_{\rm bs} n_{\rm b}/N_{\rm s}$,
by solving simultaneous linear equations,
$N_{\rm s} = n_{\rm s} + f_{\rm bs} n_{\rm b}$ and 
$N_{\rm c} = f_{\rm sc} n_{\rm s} + n_{\rm b}$, where
$N_{\rm s}$ $(N_{\rm c})$ is the number of observed events in the signal 
(control) region, and $n_{\rm s}$ $(n_{\rm b})$ is the number of the signal 
(background) events in the signal (control) region.
The background component thus obtained is 14\% of the entire candidate 
event sample at $W<1.3$~GeV. 
Above 1.3~GeV, the background is less than 1\% and is negligibly small.

\section{Derivation of the cross section}
\label{sec:dcs}

Similarly to the derivation of the $\pi^0 \pi^0$ cross section~\cite{masuda},
we first define and evaluate the $e^+e^-$-based
cross section separately for the p-tag and e-tag samples.
After confirming the consistency between the p- and e-tag measurements 
to ensure validity of the efficiency corrections, 
we combine their yields and efficiencies. 
We then convert the $e^+e^-$-incident-based differential cross section
to that based on $\gamma^* \gamma$-incident by dividing
by the single-tag two-photon luminosity function 
$d^2 L_{\gamma^* \gamma}/dW dQ^2$, which is a function of $W$ and $Q^2$. 
We use the relation
\begin{eqnarray}
%\begin{align*}
\sigma_{\rm tot}(\gamma^* \gamma \to \ks \ks) =
\frac{1}{2 \frac{d^2 L_{\gamma^* \gamma}}{dW dQ^2}} \times \nonumber \\
\frac{Y(W, Q^2)}{(1+\delta)\varepsilon(W, Q^2) \Delta W \Delta Q^2\!\int\!{\cal L}dt{\cal B}^2}, 
\label{eqn:cscomb}
%\end{align*}
\end{eqnarray}
where $Y$ is the yield and $\varepsilon$ is the efficiency obtained by the signal MC.
Here, the factor $\delta$ corresponds to the radiative correction, 
$\int {\cal L}dt$ is the integrated luminosity of 759~fb$^{-1}$, 
and ${\cal B}^2 = 0.4789$ is the square
of the decay branching fraction ${\cal B}(\ks \to \pi^+ \pi^-)$.
The measurement ranges of $W$ and $Q^2$, and the 
corresponding bin widths $\Delta W$ and 
$\Delta Q^2$, are summarized in Table~\ref{tab:bins}.
Our measurement extends down to the mass threshold
$W=2m_{\ks}$, where $m_{\ks}$ is the mass of $\ks$~\cite{pdg2016}.
For bins for $W > 1.2$~GeV,
the cross section is first calculated with $\Delta W = 0.05$~GeV, 
and then its values in two or four adjacent bins are combined, with 
the point plotted at the arithmetic mean of the entries in that combined bin.

\begin{center}
\newlength{\myheight}
\setlength{\myheight}{4mm}
\begin{table}[htb]
\caption{
The measurement range and bin widths defining the bins in the two-dimensional
 $(W, Q^2)$ space.
}
\begin{tabular}{c|cccc}
\hline \hline
Variable & Measurement & Bin width & Unit & Number \\
         & range   & & & of bins\\
\hline \hline
$W$ & 0.995($2m_{\ks}$) -- 1.05 & 0.055 & GeV & 1\\
    & 1.05 -- 1.2 & 0.05 &   & 3\\
    & 1.2 -- 1.6 & 0.1 &    & 4\\
    & 1.6 -- 2.6 & 0.2 &   &  5\\
\hline
\rule{0cm}{\myheight}$Q^2$ & 3.0 -- 7.0  & 2.0 & GeV$^2$ & 2\\
 & 7.0 -- 10.0 & 3.0 & & 1\\
 & 10.0 -- 15.0 & 5.0 & & 1\\
 & 15.0 -- 30.0 & 15.0 & & 1\\
\hline \hline
\end{tabular}
\label{tab:bins}
\end{table}
\end{center}

\subsection{Efficiency plots and consistency check for the
p-tag and e-tag measurements}
\label{sub:efplot}

\begin{figure*}
\centering
\includegraphics[width=15cm]{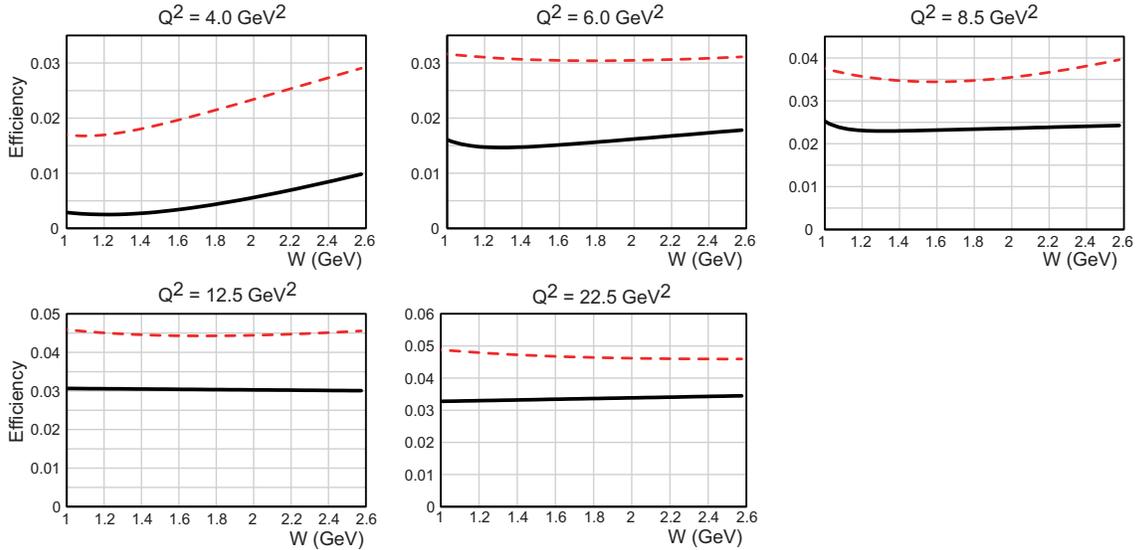}
\centering
\caption{Efficiency (including trigger effects)
as estimated from the signal-MC samples. 
The solid (black) and dashed (red)
curves are for e-tag and p-tag events, respectively.
Results are shown for five $Q^2$ regions, whose central
values are indicated above each panel.   
}
\label{fig:mceff}
\end{figure*}

Figure~\ref{fig:mceff} shows the aggregate efficiencies, 
as a function of $W$ for
the selected $Q^2$ bins of the p- or e-tag samples,
including all event selection and trigger effects.
These efficiencies are obtained from the signal-MC events, which are 
generated assuming an isotropic $\ks$ angular distribution
in the $\gamma^* \gamma$ c.m. frame. 

\begin{figure*}
\centering
\includegraphics[width=15cm]{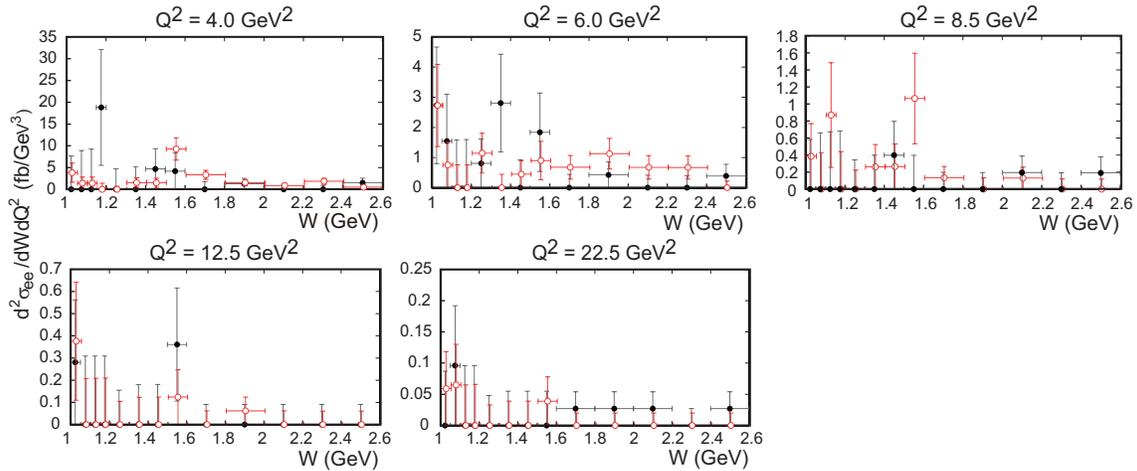}
\centering
\caption{The efficiency-corrected and
background-subtracted $W$ dependence of the $e^+e^-$-based cross section
in each $Q^2$ bin.
The black closed (red open) circles with error bars  are for the 
e-tag (p-tag) measurements. 
The red p-tag points have been shifted slightly to the right for enhanced visibility. 
}
\label{fig:cs_tags}
\end{figure*}

Our accelerator and detector systems are asymmetric between
the positron and electron incident directions and energies, and separate
measurements of the p-tag and e-tag samples provide a good
internal consistency check for various systematic effects of
the trigger, detector acceptance, and selection conditions.
Figure~\ref{fig:cs_tags} compares the $e^+e^-$-based cross section measured 
separately for the p- and e-tags.
They are expected to show the same cross section 
according to the $C$ symmetry if there is no systematic bias.
In this figure, the estimated non-exclusive backgrounds are subtracted,
fixing the ratio of the values from the p- and e-tag measurements.

The results from the two tag conditions are consistent within 
statistical errors.
We therefore combine the p- and e-tag sample results using their summed
yields and averaged efficiencies.

\subsection{Derivation of angle-integrated \boldmath{$\gamma^* \gamma \to \ks \ks$} cross section}
\label{sub:radcor}
We apply a radiative correction of 2\% 
to the total cross section.
This value is the same as that evaluated in the analogous case of single pion 
production~\cite{pi0tff}.
This correction depends only slightly on $W$ and $Q^2$, 
and is treated as a constant.
The radiative effect in the event topology is taken into account in the signal-MC event generation 
and is reflected in the efficiency calculation.

To account for the non-linear dependence on $Q^2$,
we define the nominal $Q^2$ for each finite-width bin
$\overline{Q}^2$, using the formula
\begin{equation}
\frac{d\sigma_{ee}}{dQ^2}(\overline{Q}^2) = \frac{1}{\Delta Q^2}
\int_{\rm bin} \frac{d\sigma_{ee}}{dQ^2}(Q^2) dQ^2 ,
\label{eqn:q2form}
\end{equation}
where $\Delta Q^2$ is the bin width.
We assume an approximate dependence
of $d\sigma/dQ^2 \propto Q^{-7}$ for this calculation~\cite{pi0tff},
independent of $W$. 
The $\overline{Q}^2$ values thus obtained are listed in 
Table~\ref{tab:repreq2}. 
We use the luminosity function at a given $\overline{Q}^2$ point to
obtain the $\gamma^* \gamma$-based cross section for each $Q^2$ bin.
We also list the central value of the $Q^2$ bins; these are used for convenience 
to represent the individual bins in tables and figures. 

\begin{center}
\begin{table}
\setlength{\myheight}{4mm}
\caption{The nominal $Q^2$ value ($\overline{Q}^2$) for each 
$Q^2$ bin.}
\begin{tabular}{cc|c}
\hline \hline
\rule{0cm}{\myheight}$Q^2$ bin (GeV$^2$) & Bin center (GeV$^2$) & $\overline{Q}^2$ (GeV$^2$) \\
\hline \hline
2 -- 3 & 2.5 & 2.42 \\
3 -- 5 & 4.0 & 3.81 \\
5 -- 7 & 6.0 & 5.87 \\
7 -- 10 & 8.5 & 8.30 \\
10 -- 15 & 12.5 & 12.1 \\
15 -- 30 & 22.5 & 20.6 \\
\hline \hline
\end{tabular}
\label{tab:repreq2}
\end{table}
\end{center}

 The $Q^2$ value measured for each event can differ from the true
$Q^2$ for two primary reasons:
the finite resolution in our $Q^2$ determination
and/or the reduction of the incident electron energy due to initial-state radiation (ISR).
However, the relative $Q^2$ resolution in the measurement, typically 0.7\%,
which is estimated using the signal-MC events, 
is much smaller than the typical bin sizes and therefore 
has a negligible effect. 
The ISR effect is also negligibly small in this analysis
owing to the tight $E_{\rm ratio}$ selection
criterion, which rejects events with high-energy radiation.
Thus, we do not apply the $Q^2$-unfolding procedure in this analysis,
which was applied in the previous analysis where the corresponding 
selection condition was less restrictive~\cite{masuda}.

\begin{figure}
\centering
\includegraphics[width=8cm]{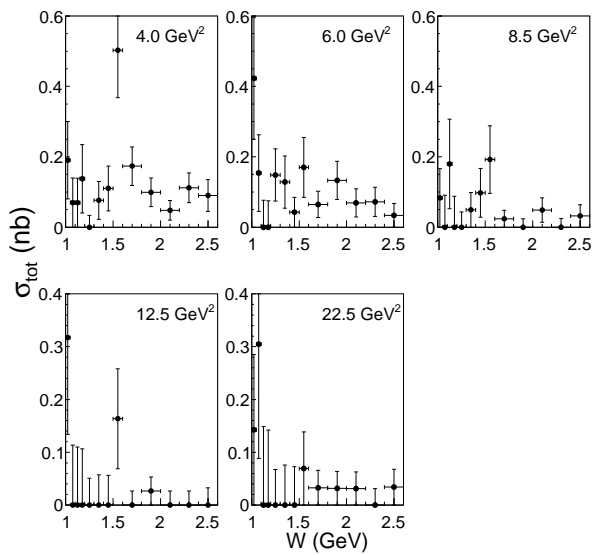}
\caption{
Total cross sections (integrated over angle)
for $\gamma^* \gamma \to \ks \ks$ 
in the five $Q^2$ bins indicated in each panel.
}
\label{fig:tot}
\end{figure}

The $e^+e^-$-based differential cross sections thus measured
are converted to $\gamma^*\gamma$-based cross sections, 
corresponding to 
$\sigma_{\rm tot}(\gamma^* \gamma \to \ks \ks) = \sigma_{\rm TT}~+~\epsilon_0 \sigma_{\rm LT}$,
using the luminosity function as described above.
Figure~\ref{fig:tot} shows the total cross sections (integrated over angle)
for the single-tag two-photon production of $\ks \ks$, 
as a function of $W$ in five $Q^2$ bins.

\subsection{
Helicity components and angular dependence
}
\label{sub:siglt}
We now estimate $\epsilon_0$ and $\epsilon_1$, the factors that appear in 
Eqs.~(\ref{eqn:t0}) -- (\ref{eqn:t2}), 
in each $(W, Q^2)$ bin.
We use the mean value of $\epsilon_0$ ($\epsilon_1$) 
as calculated by Eq.~(\ref{eqn:eps0}) [Eq.~(\ref{eqn:eps1})]
for the selected events from the signal-MC samples, as they depend only 
very weakly on $Q^2$ and $W$. 
The numerical values {in the kinematic range $W < 1.8\,{\rm GeV}$} are summarized in Table~\ref{tab:eps0},
where we neglect the $W$ dependence because
it is small (within $\pm 2\%$); here, 
we apply the partial-wave analysis of Sec. VI.

For analysis of the three helicity components 0, 1, and 2
described in Sec.~\ref{sub:tff},
we use a normalized angular-differentiated cross section 
(integrated over $Q^2$)
$(d^2 \sigma/d |\cos \theta^*| d |\varphi^*|)/\sigma$,
which is derived as follows.
We assume that the angular dependence of 
$d^2 \sigma/d |\cos \theta^*| d |\varphi^*|$ follows 
$N_{\rm EXP}(|\cos \theta^*|, |\varphi^*|)/N_{\rm MC}(|\cos \theta^*|, |\varphi^*|)$
in each $W$ bin integrated in the $Q^2 = $~3 -- 30~GeV$^2$ region and
take this to be the angular dependence at 
$Q^2 = \langle Q^2 \rangle = 6.5$~GeV$^2$, 
where $\langle Q^2 \rangle$ is the mean 
value of $Q^2$ for all the selected experimental events. 
For this purpose, we use four $W$ bins starting at the mass threshold:
0.995~ -- 1.2~GeV, 1.2 -- 1.4~GeV, 1.4 -- 1.6~GeV,
and 1.6 -- 1.8~GeV. 
The angular bin sizes are $\Delta |\cos \theta^*|=0.2$ and 
$\Delta |\varphi^*| = 30^\circ$.
We use the normalization 
$ \int_0^1 d |\cos \theta^*| \int_0^\pi d |\varphi^*| 
[(d^2 \sigma/d |\cos \theta^*| d |\varphi^*|)/\sigma] = 1$.

\begin{center}
\begin{table}
\setlength{\myheight}{4mm}
\caption{
The values of the $\epsilon_0$ and $\epsilon_1$ parameters, as a function of $Q^2$,
at $W < 1.8$~GeV.
}
\begin{tabular}{c|cc}
\hline \hline
\rule{0cm}{\myheight} $Q^2$ bin (GeV$^2$) &  $\epsilon_0$ &  $\epsilon_1$ \\
\hline \hline
3 -- 5 & 0.92 & 1.33 \\
5 -- 7 & 0.91 & 1.32 \\
7 -- 10 & 0.89 & 1.30  \\
10 -- 15 & 0.87 & 1.28 \\
15 -- 30 & 0.82 & 1.23 \\
\hline \hline
\end{tabular}
\label{tab:eps0}
\end{table}
\end{center}

\subsection{Derivation of the partial decay width of the \boldmath{$\chi_{cJ}$} mesons}
We find a clear excess of events in the mass region of the $\chi_{cJ}$ ($J=0,2$) mesons as
shown in Fig.~\ref{fig:wdist}. 
We define signal regions to be 3.365 -- 3.465~GeV/$c^2$ and 
3.505 -- 3.605~GeV/$c^2$ for the $\chi_{c0}$ and $\chi_{c2}$ 
mesons, respectively, and 
note that the process $\chi_{c1} \to \ks\ks$ is prohibited by parity conservation.
We measure over the range 2~GeV$^2 \le Q^2 \le 30$~GeV$^2$, 
and expect a much better efficiency in the $\chi_{cJ}$ mass region
at small $Q^2$ than in the lower-$W$ region.

The charmonium yields in the $Q^2$ range are 7 and 3 for the $\chi_{c0}$ and $\chi_{c2}$ mesons,
respectively; we assume, given the evident absence of background, that 
they are pure contributions from charmonia. 
Based on studies of no-tag $\ks\ks$~\cite{ksks} and single-tag 
$\pi^0\pi^0$~\cite{masuda} measurements, we similarly estimate
less than one background event for the total of the two 
charmonium regions.

We first determine the $e^+e^-$-based cross section in the
two $\chi_{cJ}$ mass regions. This is then translated to the product 
of the two-photon decay width and the branching fraction into the $\ks\ks$
final state using the relation 
\begin{eqnarray}
\frac{d\sigma_{ee}}{dQ^2} &=&  
4\pi^2 \left(1+\frac{Q^2}{M_R^2}\right)\frac{(2J+1)}{M_R^2}
\frac{2 d^2 L_{\gamma^* \gamma}}{dW dQ^2}  \nonumber \\
& &  \times\, \Gamma_{\gamma^*\gamma}(Q^2){\cal B}(\ks\ks),
\end{eqnarray}
which is valid for a narrow resonance after integrating over $W$, 
where $M_R$ is the resonance mass.  
It is not possible to present the $\chi_{cJ}$ production
rate as a function of $\sigma_{\gamma^* \gamma} (W, Q^2)$
because we know that each of the  $\chi_{cJ}$ mesons has a narrow
but finite width that is comparable to 
the resolution of our measurement. 
Instead, we present the two-photon decay width $\Gamma_{\gamma^*\gamma}(Q^2)$ with the above formula,
which we define similarly to the TFF in Eq.~(\ref{eqn:arj})
with respect to the functional dependence on $Q^2$.

Note that the three independent helicity amplitudes are effectively added 
in this definition, assuming unpolarized $e^+e^-$ collisions 
for the $\chi_{c2}$ meson, and this formula can be considered as the definition 
of $\Gamma_{\gamma^*\gamma}(Q^2)$ at $Q^2>0$; we adopt it as such in what follows.

\begin{figure}
 \centering
   {\epsfig{file=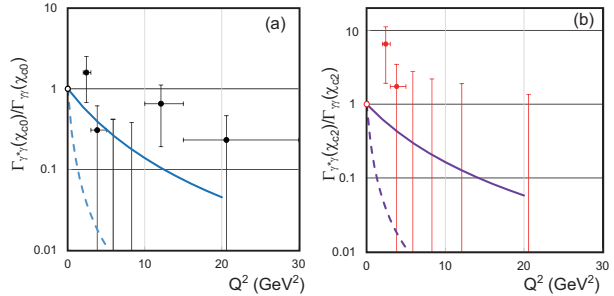,width=80mm}}
 \caption{$Q^2$ dependence of $\Gamma_{\gamma^*\gamma}$ for the
(a) $\chi_{c0}$ and (b) $\chi_{c2}$ mesons normalized to $\Gamma_{\gamma\gamma}$
(at $Q^2=0$)~\cite{pdg2016}. The data point without a dot is based on a 
zero-event observation, and the upper edge of its error bar corresponds to
the value for one event.
The overall uncertainties due to the normalization errors of 
the $\Gamma_{\gamma\gamma} {\cal B}(\ks \ks)$ are not shown.
The solid and dashed curves, respectively, show the SBG~\cite{schuler} prediction
and also one motivated by VDM, assuming $\rho$ dominance.
}
\label{fig:charmonium_ggg}
\end{figure}

Figure~\ref{fig:charmonium_ggg} shows the $Q^2$ dependence of 
$\Gamma_{\gamma^*\gamma}/\Gamma_{\gamma\gamma}$ for the $\chi_{c0}$
and $\chi_{c2}$ mesons, where $\Gamma_{\gamma\gamma}$ is the value for
the real two-photon decay, which is extracted from the 
$\Gamma_{\gamma\gamma}{\cal B}(\ks \ks)$ world-average values of ($7.3~\pm~0.6$)~eV and ($0.291~\pm~0.025$)~eV
for the $\chi_{c0}$ and $\chi_{c2}$ mesons, respectively~\cite{pdg2016}. 
This is the first measurement of $\chi_{cJ}$ charmonium production
in high-$Q^2$ single-tag two-photon collisions.

These measurements are compared to the SBG~\cite{schuler} predictions
evaluated at the 
$\chi_{cJ}$ mass and also the expectation 
using a vector-dominance model (VDM)~\cite{sakurai} 
with the $\rho$ mass in the factor
$(1+Q^2/m_\rho^2)^{-2}$. As can be clearly seen, the low statistics notwithstanding,
we obtain reasonable agreement with SBG prediction 
at the charmonium-mass scale.  

\subsection{Systematic uncertainties}
\label{sub:sysunc}
We estimate systematic uncertainties in the measurement of 
the differential cross section as summarized in Table~\ref{tab:syserr}.

\subsubsection{Uncertainties in the efficiency evaluation}
The detection efficiency is evaluated using signal-MC events.
However, our simulation has some known mismatches with data that 
translates into uncertainties in the efficiency evaluation.

Charged particle tracking has a 2\% uncertainty for five tracks, 
which is estimated from a study of the decays 
$D^{*\pm} \to D^0 \pi^{\pm}, D^0 \to K^0_S (\to \pi^+ \pi^-) \pi^+ \pi^-$
(0.35\% per track) including an uncertainty in the radiation
by an electron within the CDC volume (about 1\%, added in quadrature). 

The electron identification efficiency in this measurement is very high, 
around 98\%, and a 1\% systematic uncertainty is assigned to it.
Detection of the $\pi^+\pi^-$ pairs for reconstructing 
two $\ks$ mesons has a 2\% uncertainty due to the requirement to identify four charged 
pions, and another 3\% for $\ks$ reconstruction 
and selection dominated by a possible 
difference in the mass resolution for the reconstructed $\ks$ 
between the experiment and the signal MC.

Our kinematic condition based on the $E_{\rm ratio}$ and $p_t$ balance
has an accompanying uncertainty of 4\%. 
In addition, imperfections in modeling detector edge locations and other 
geometrical-description effects result in an uncertainty of 1\%.

The uncertainty of the trigger efficiency is 
estimated using different
types of subtrigger components, with special attention given to events satisfying 
multiple trigger conditions.
We select four kinds of primary subtriggers whose efficiencies are well-studied. 
The first two are distinct possible two track triggers: one requires
total energy activity in the ECL exceeding 0.5 GeV, and the other
requires an ECL cluster
as well as two TOF hits. The other two trigger lines are the neutral triggers,
namely HiE
and Clst4. 

More than half of the signal 
candidates are triggered by two or more distinct triggers.
We estimate the uncertainty on the trigger inefficiency as a 
fractional difference of the efficiencies between the 
cases for which all the subtrigger components are ORed and the case where 
at least one of the selected four triggers is fired.
This uncertainty is estimated to be 3\% for $W<2.6~\GeV$ and 1\% for the $\chi_{cJ}$ 
charmonium-mass region. 

Backgrounds overlapping with the signal events may reduce 
the efficiency; this effect is accounted for in MC simulations by embedding 
hits from a non-triggered event (``random'' or ``unbiased'' triggers) in each signal-MC event. 
We evaluate this effect separately for each different
beam-condition state and run period.  
The corresponding effect on the efficiency is estimated to be 2\%. 

We take into account an uncertainty on the efficiency-correction 
factor arising from the angular dependence of the differential cross section. 
This correction arises when both the selection efficiency and differential 
cross sections have angular nonuniformities.  
As we do not measure the angular dependence of the differential cross
section for different kinematic regions owing to limited
statistics, we assume several typical angular dependences
of the differential cross section 
based on the spherical-harmonic functions of 
$J \leq 2$: proportional to 
$\cos \theta^*$, $\cos^2 \theta^*$,
$(3 \cos^2 \theta^*-1)^2$, $\sin^4 \theta^*$, 
$(1+ 0.5 \cos \varphi^*)$, and $(1+ 0.5 \cos 2 \varphi^*)$. 

We examine the efficiency differences for these angular-dependence shapes 
from that of the isotropic-efficiency case using simulated events,
and assign its typical variation size, taking a quadratic sum
of the $\cos \theta^*$ and $\varphi^*$ contributions,
to the systematic uncertainty from this source. 
The $W$-dependent estimated error magnitude is 6\% -- 22\%:
this dependence originates purely from the difference in the 
degree of nonuniformity in the efficiency.

\subsubsection{
Uncertainties from other sources
}

We assign 7\%, half of the magnitude of the
subtraction itself, as the uncertainty in the 
background subtractions arising from $\ks \ks X$ non-exclusive 
processes for $W<1.3~$GeV. 
We assign 3\% as the uncertainty for the other $W$ regions. 
Other background sources are negligibly small.

The omission of the $Q^2$-unfolding procedure introduces an uncertainty of 1\%.
The radiative correction has an uncertainty of 3\%. 
The evaluation of the luminosity function gives an uncertainty of 4\%, 
including a model uncertainty for the form factor of 
the untagged side (2\%)~\cite{masuda}.
The integrated luminosity measurement has an uncertainty of 1.4\%.

The systematic uncertainties are added in quadrature unless
noted above. 
The total systematic uncertainty is between 13\% and 24\%,
depending on the $W$ bins.

\begin{center}
\begin{table}
\caption{Sources of systematic uncertainties. The values are
indicated for specific $W$ ranges.
DCS stands for the differential cross section.}
\label{tab:csys}
\begin{tabular}{c|c}
\hline \hline
Source & Uncertainty (\%) \\
\hline \hline
Tracking & 2 \\
Electron-ID & 1 \\
Pion-ID (for four pions) & 2 \\
$\ks$ reconstruction (for two $\ks$'s) & 3 \\
Kinematic selection & 4 \\
Geometrical acceptance & 1 \\
Trigger efficiency & 1 -- 3 \\
Background effect for the efficiency & 2 \\
Angular dependence of DCS & 6 -- 22 \\
Background subtraction & 3 -- 7 \\
No unfolding applied & 1 \\
Radiative correction & 3 \\
Luminosity function & 4\\
Integrated luminosity & 1.4\\
\hline
Total & 13 -- 24 \\
\hline \hline
\end{tabular}
\label{tab:syserr}
\end{table}
\end{center}

\section{Measurement of the transition form factor}
\label{sec:meastff}
In the measurement of the no-tag mode of the process 
$\gamma \gamma \to \ks \ks$~\cite{ksks},
the $f_2'(1525)$ resonance with a structure corresponding to the 
$f_2(1270)$ and the $a_2(1320)$ mesons, and their destructive interference, were observed.

In the present single-tag measurement (Fig.~\ref{fig:tot}),
a structure corresponding to the $f_2'(1525)$ state is clearly visible.
A structure near the threshold of $\ks \ks$ is also visible that 
may be associated with the $f_0(980)$ and the $a_0(980)$ mesons.
We do not find any prominent enhancement at the 
$f_2(1270)$ or the $a_2(1320)$ mass, and this feature is consistent with
destructive interference. 

In this section, we extract the $Q^2$ dependence of the helicity-0, -1, and -2 
TFF of the $f_2'(1525)$ meson and compare it with theory.
We also compare the $Q^2$ dependence of cross sections near the threshold 
with theory.

\subsection{Partial wave amplitudes}
\label{sub:pwa}
The helicity amplitudes in Eq.~(\ref{eqn:dsdcos}) can be written
in terms of S and D waves in the energy region $W \leq 1.8~\GeV$,
identical to the expressions presented in our similar study of 
$\pi^0 \pi^0$ production~\cite{masuda}.
For completeness, we reproduce here the expression of the $t_0$, $t_1$, and $t_2$ amplitudes in 
Eqs.~(\ref{eqn:t0}) to (\ref{eqn:t2}) in terms of S and D waves:
\begin{eqnarray}
t_0 &=& |S Y^0_0 + D_0 Y^0_2|^2 + |D_2 Y^2_2|^2 
  + 2 \epsilon_0 |D_1 Y^1_2|^2 , \nonumber \\
t_1 &=& 2 \epsilon_1 \Re \left[ (D_2^* |Y^2_2| - S^* Y^0_0 - D_0^* Y^0_2)
D_1 |Y^1_2| \right], \nonumber \\
t_2 &=& -2 \epsilon_0 \Re \left[D_2^* |Y^2_2| (S Y^0_0 + D_0 Y^0_2) \right] ,
\label{eqn:pwtheta}
\end{eqnarray}
where $S$ is the S-wave amplitude,
$D_0$,\; $D_1$, and $D_2$ denote the helicity-0, -1, and -2 components
of the D wave, respectively,~\cite{pw} 
and $Y^m_J$ are the spherical harmonics.
We use the absolute values for the spherical
harmonics since the helicity amplitudes are independent of
$\varphi^*$~\cite{gss}.

After integrating over the azimuthal angle, the differential cross section
can be expressed as:
\begin{eqnarray}
 \frac{d \sigma (\gamma^* \gamma \to \ks \ks)}
{4 \pi d |\cos \theta^*|} 
= \left| S \: Y^0_0 + D_0 \: Y^0_2 \right|^2 &&  
\nonumber \\
 + 2 \epsilon_0 \left|D_1 \: Y^1_2 \right|^2  
+ \left|D_2 \: Y^2_2 \right|^2 .&&
\label{eqn:dsdc}
\end{eqnarray}

The angular dependence of the cross section is contained in the 
spherical harmonics, while the $W$ and $Q^2$ dependences are 
determined by the partial waves.
The $Q^2$ dependence is governed by the transition form factors of the 
resonances
and the helicity fractions in D waves. 
The $W$ dependence is expressed by the relativistic Breit-Wigner function
and the energy dependence of the non-resonant backgrounds. 

\subsection{Parameterization of amplitudes}
\label{sub:param}
We extract the $Q^2$ dependence of $F_{f2p}(Q^2)$, the TFF of the $f_2'(1525)$ meson,
by parameterizing $S$, $D_0$, $D_1$, and $D_2$
and fitting the event distribution in the energy region 
$1.0~\GeV \le W \le 1.8~\GeV$.

Both isoscalar $f$ and isovector $a$ mesons 
contribute to two-photon production of a $\ks$ pair.
The relative phase between the $f_2(1270)$ and the $a_2(1320)$ mesons was
found to be fully destructive in the previous
no-tag measurement of this process~\cite{ksks}.
Correspondingly, we assume the phase to be $180^\circ$, independent of $Q^2$.

The partial-wave amplitudes $S$ and $D_i \; (i=0,1,2)$ are parameterized as follows:
\begin{eqnarray}
S &=& A_{BW} e^{i \phi_{BW}}~+~ B_S e^{i \phi_{BS}}  , \nonumber \\
D_i &=& \sqrt{r_{ifa}(Q^2)} (A_{f_2(1270)}-A_{a_2(1320)})e^{i \phi_{faDi}} \nonumber \\
&& + \sqrt{r_{ifp}(Q^2)} A_{f_2'(1525)}e^{i \phi_{fpDi}} \nonumber \\
&& + B_{Di} e^{i \phi_{BDi}}, 
\label{eqn:param}
\end{eqnarray}
where 
$A_{f_2(1270)}$, $A_{a_2(1320)}$, and $A_{f_2'(1525)}$ 
are the amplitudes of the $f_2(1270)$, the  $a_2(1320)$, and the $f_2'(1525)$
mesons, respectively,
and $A_{BW}$ is an S-wave amplitude, as explained below.
The parameters $r_{ifa}(Q^2)$ and $r_{ifp}(Q^2)$ designate the fractions of
the $f_2(1270)/a_2(1320)$ 
and the $f_2'(1525)$-contribution in the D$_i$ wave, respectively, 
with the unitarity constraint of $r_{0j} + r_{1j} + r_{2j} = 1$, and
$r_{ij} \ge 0$, where $j$ stands for $fa$ or $fp$.
$B_S$ and $B_{Di}$ are nonresonant ``background'' amplitudes 
for S and D$_i$ waves;
$\phi_{BS}$, $\phi_{BDi}$, $\phi_{BW}$, and $\phi_{jDi}$
are the phases of these S-wave and D$_i$-wave background amplitudes, 
of the amplitude $A_{BW}$, and of the amplitudes of the $f_2(1270)/a_2(1320)$ and the $f_2'(1525)$-contribution in D$_i$ wave;
they are assumed to be independent of $Q^2$ and $W$.
The overall arbitrary phase is fixed by taking $\phi_{fiD0} = 0$.

Here, we describe the parameterization of the $f_2(1270)$, the $a_2(1320)$, and
the $f_2'(1525)$ mesons.
The relativistic Breit-Wigner resonance amplitude
$A_R^J(W)$ for a spin-$J$ resonance $R$ of mass $m_R$ is given by
\begin{eqnarray}
A_R^J(W) &=& F_R(Q^2) \sqrt{1 + \frac{Q^2}{m_{R}^2}} 
\sqrt{\frac{8 \pi (2J+1) m_R}{W}} 
\nonumber \\
&& \times \frac{\sqrt{ \Gamma_{\rm tot}(W)
\Gamma_{\gamma \gamma}(W) \B(\ks \ks)}}
{m_R^2 - W^2 - i m_R \Gamma_{\rm tot}(W)} \; ,
\label{eqn:arj}
\end{eqnarray}
where $F_R(Q^2)$ is the TFF of the resonance $R$, and is
defined by the above formula in relation to the
tagged two-photon cross section~\cite{masuda} (see also Eq. (C13) and 
(C28) in Ref~\cite{ppv}).
The energy-dependent total width $\Gamma_{\rm tot}(W)$ is given by
Eq.~(38) in Ref~{\cite{masuda}}.

Since the TFF and the fractions of the $f_2(1270)$ meson have been 
measured~\cite{masuda},
we accordingly fit the data with a smooth function of $Q^2$. 
We have used the obtained functions for Eq.~(\ref{eqn:arj}), \textit{viz.} 
$F_{f2}(Q^2) = 1/(1+3.3 \times Q^2)^{0.94}$,
$r_{0fa}(Q^2) = 0.015 \times Q^2 + 0.30$, and 
$r_{1fa}(Q^2) = 0.15 \times (Q^2/9.6)^{-0.2}$, with $Q^2$ in GeV$^2$. 
Since the $a_2(1320)$ and the $f_2(1270)$ mesons are so close in mass, 
we assume they have identical TFFs.

In the $\gamma\gamma \rightarrow \ks\ks$ reaction, 
a peak structure near the threshold is predicted
even though a destructive interference between the $f_0(980)$ and the $a_0(980)$ 
states is expected to suppress such events~\cite{Pennington}.
Thus, we employ a Breit-Wigner function 
or a power-law function, shown in the first line of Eq.~(\ref{eqn:para2})
in the description of the S wave. 
In the case of the Breit-Wigner function, the amplitude $A_{BW}$ is 
parameterized as
\begin{eqnarray}
A_{BW}(W) &=& 
\sqrt{\frac{8 \pi  m_S}{W}} 
\frac{f_S}{m_S^2 - W^2 - i m_S g_S}  
\nonumber \\
&& \times \frac{1}{(Q^2/m_0^2 + 1)^{p_S}},
\label{eqn:bw_s}
\end{eqnarray}
where $m_S$ is the mass of the resonance,
$f_S$ parameterizes the amplitude size, and
$g_S$ is the total width of the resonance.
We assume a power-law behavior for the $Q^2$ dependence,
where $p_S$ is the power. 
We take $m_S=0.995~\GeV/c^2$ by assuming that the resonance coincides with
the $K \bar{K}$ threshold.

We assume a power-law behavior in $W$ for the background amplitudes, which are then
multiplied by the threshold factor $\beta^{2 \ell +1}$ 
(with $\ell$ denoting the orbital angular momentum of the two-$\ks$ system), and 
with an assumed $Q^2$ dependence for all the waves:
\begin{eqnarray}
B_S &=&  \frac{\beta a_S \left( W_0/W \right)^{b_S}}
{(Q^2/m_0^2 + 1)^{c_S}}
\;, \nonumber \\
B_{D0} &=& \frac{\beta^5 a_{D0} \left( W_0/W \right)^{b_{D0}}}
{(Q^2/m_0^2 + 1)^{c_{D0}}} \;, \nonumber \\
B_{D1} &=& \frac{\beta^5 Q^2 a_{D1} \left( W_0/W \right)^{b_{D1}}}
{(Q^2/m_0^2 + 1)^{c_{D1}}} \;, \nonumber \\
B_{D2} &=& \frac{\beta^5 a_{D2} \left( W_0/W \right)^{b_{D2}}}
{(Q^2/m_0^2 + 1)^{c_{D2}}} \;, 
\label{eqn:para2}
\end{eqnarray}
where $\beta=\sqrt{1 - 4m_{\ks}^2/W^2}$ is the $\ks$ velocity 
divided by the speed of light. 
We take $W_0=1.4~\GeV$ and $m_0=1.0~\GeV /c^2$.
Note that $B_{D1}$ has an additional factor of $Q^2$ to ensure that
this amplitude vanishes at $Q^2=0$.
We set $a_i \ge 0 \; (i=S, D_0, D_1, D_2)$ to fix the arbitrary sign 
of each background amplitude,
thereby absorbing the sign into the corresponding phase. 

All parameters of the $f_2(1270)$, the $a_2(1320)$, and the $f_2'(1525)$
mesons are fixed at the PDG values~\cite{pdg2016}.
The normalization of the TFF is such that $F_{f2p}(0)=1.00 \pm 0.07$;
the error reflects the uncertainty of its two-photon decay
width at $Q^2 = 0$~\cite{pdg2016}.

\subsection{Extracting the TFF of the \boldmath{$f_2'(1525)$} meson}
\label{sub:tff}
We employ a partial wave analysis to extract the TFF of the $f_2'(1525)$
meson separately for helicity=0, 1, and 2, realizing that there is a fundamental limitation 
due to the inherent correlation in $S$, $D_0$, $D_1$, and $D_2$~\cite{masuda}.
To overcome this limitation, we simultaneously fit both 
the $Q^2$-integrated differential cross sections
and the total cross section.
The former is a function of $W$, $|\cos \theta^*|$, and $|\varphi^*|$ while 
the latter is a function of $W$ and $Q^2$.

The $Q^2$-integrated differential cross sections 
are divided into six $|\varphi^*|$ bins, of equal $30^\circ$ width,
five $|\cos \theta^*|$ bins with a bin width of 0.2,
and five $W$ bins covering 1.0 -- 1.2~GeV, 1.2 -- 1.4~GeV, 1.4 -- 1.6~GeV, 
1.6 -- 1.8~GeV, and 1.8 -- 2.6~GeV.
The average value of $Q^2$, $\langle Q^2 \rangle$, is 6.5~GeV$^2$.

The $Q^2$-integrated differential cross sections together with 
the total cross sections
are fitted with the parameterization described above.
In the fit, the usual $\chi^2$ is replaced by $\chi^2_{\rm P}$ with its equivalent 
Poisson-likelihood quantity $\lambda$ defined in Ref.~\cite{BC}: 
\begin{equation}
\chi^2_{\rm P} \equiv -2 \ln \lambda 
= 2 \sum_i \left[ p_i - n_i + n_i \ln \left( \frac{n_i}{p_i} 
\right) \right] ,
\label{eqn:plike}
\end{equation}
where $n_i$ and $p_i$ are the numbers of events observed
and predicted in the $i$-th bin and the sum is over all bins.

We minimize the sum of two $\chi^2_{\rm P}$ values for the
$Q^2$-integrated differential and total cross sections:
\begin{equation}
\chi^2_{\rm comb} = \chi^2_{\rm P}(W,|\cos \theta^*|,|\varphi^*|) 
+ \chi^2_{\rm P} (W,Q^2).
\label{eqn:chisum}
\end{equation}

In the first term,
the predicted number of events in each $W$ bin
is normalized such that the differential cross section integrated over 
$|\cos \theta^*|$ and $|\varphi^*|$ is equal to the total cross section 
in each $W$ bin.
In the second term, the predicted cross section value
is converted to the number of events by multiplying by a known 
conversion factor.
These two subsets of data are obtained from the same data sample, but the 
correlation between the two is negligible. 
The effect of limited statistics in using this combined $\chi^2_{\rm P}$ is negligible
since the $Q^2$-integrated differential cross sections and the total cross sections are 
almost independent.
We float the normalization factors in the $Q^2$-integrated differential 
cross sections and fix them in the total cross sections  
so as to minimize the correlation between the two sets of data in the fit.

Here, we include zero-event bins in calculating the $\chi^2_{\rm P}$ given in 
Eq.~(\ref{eqn:plike}).
In fitting using Eq.~(\ref{eqn:plike}), systematic uncertainties on the cross 
section are not taken into account. 
Their effects are detailed separately in 
Sec.~\ref{sub:systff}. 

The TFFs for the $f_2'(1525)$ meson are floated in each $Q^2$ bin,
while $r_{0fp}(Q^2)$, $r_{1fp}(Q^2)$, and $r_{2fp}(Q^2)$ are assumed such that 
\begin{equation}
r_{0fp}:r_{1fp}:r_{2fp} = k_0 Q^2 : k_1 \sqrt{Q^2} : 1 \: ,
\label{eqn:r012}
\end{equation}
where the parameters $k_0$ and $k_1$ are floated.
This parameterization is motivated by SBG~\cite{schuler} (Table~\ref{tab:pred})
 and reproduces well the measured data on the 
$f_{2}(1270)$ meson~\cite{masuda}.

In this procedure, three categories of fits are conducted:
category 1 ($A_{BW}\ne 0 \ \bigcap\ B_S=0 $),
category 2 ($A_{BW}=0 \ \bigcap\ B_S \ne 0 $),
and category 3 ($A_{BW}=B_{S}=0$).
We have assumed that the 
S wave is described only with a Breit-Wigner function 
in category 1 
and a power-law behavior in $W$ in category 2. 
The S wave is assumed not to be present in category 3.
We have also assumed $B_{D0}=B_{D1}=B_{D2}=0$ in all cases, and later assess the systematic errors
associated with this assumption.
In each category, we fit the data under the condition that either $k_0$ 
and $k_1$ are both floated, or one is floated with the other magnitude set to zero.

In category 1, the condition $k_0\ne0 \ \bigcap\ k_1\ne0$ admits two 
solutions with $\chi^2/ndf$ of 152.4/150 and 159.8/150, respectively, where
$ndf$ is the number of degrees of freedom in the fit.
Because they are smaller than the value of 173.1/151 obtained by setting $k_0=0$,
or 166.4/151 obtained by setting $k_1=0$, 
only the two solutions corresponding to $k_0\ne0 \ \bigcap\ k_1\ne0$  
are shown in Table~\ref{tab:comfit}; these are denoted as solution 1a and 1b. 
In category 2, the condition of $k_0\ne0 \ \bigcap\  k_1\ne0$ gives 
two solutions with $\chi^2/ndf$ values of 154.9/151 (solution 2a) and 156.1/151 (solution 2b), respectively.
Here again, setting $k_0=0$ or $k_1=0$ give a much larger $\chi^2$ value.
In category 3, only the solution giving the minimum $\chi^2 $ for
$k_0\ne0 \ \bigcap\ k_1 \ne0$ is listed in Table~\ref{tab:comfit}.

These fit results show that there is a significant helicity-0 component of the 
$f_2'(1525)$ meson in two-photon production when one of the photons is 
highly virtual, and also favor a non-zero
helicity-1 component of the $f_2'(1525)$ meson.
One of the solutions of a Breit-Wigner model for the S wave gives the global-minimal $\chi^2$;
nevertheless, we cannot conclude definitively that the threshold enhancement is of the Breit-Wigner type.

To extract each helicity component of the $f_2'(1525)$ meson, 
we use the values of $k_0$, $k_1$, and the TFF of the 
$f_2'(1525)$ meson that best match our data.
Both solutions (1a and 1b) in category 1 
with $k_0\ne0 \ \bigcap\ k_1\ne0$ are shown in Table~\ref{tab:comfit}.
Solutions 1a and 1b give only slight differences in their fitted values,
except for the phases $\phi_{fiD1}$ (which are opposite one another)
and solution 1a gives 7.4 smaller units of $\chi^2$ than solution 1b.
Solutions 2a and 2b are identical to solution 1a within errors  
except for the phases $\phi_{fiD1}$,  
and give 2.5 and 3.7 larger units of $\chi^2$ than solution 1a, respectively.
Thus, we take solution 1a as the nominal fit result instead of combining 
these solutions statistically.

\begin{figure}
 \centering
  {\epsfig{file=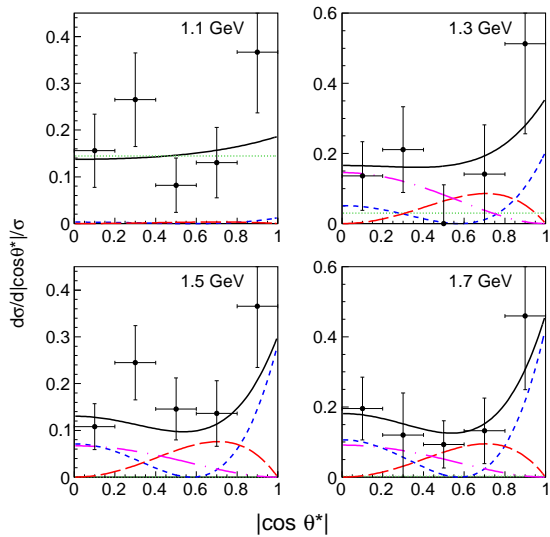,width=77mm}}
 \caption{
$|\cos \theta^*|$ dependence of the normalized differential cross sections and 
the fitted results in the four $W$ bins indicated in each panel. 
The lines shown are obtained from the nominal fit 
(at $\langle Q^2 \rangle = 6.5~\GeV^2$).
Black solid lines show the total, 
green dotted the $|S|^2$ term, blue dashed the $|D_0|^2$ term,
red long-dashed the $|D_1|^2$ term, and magenta dash-dotted the $|D_2|^2$ term.
}
 \label{fig:dcos_w1}
\end{figure}

Figure \ref{fig:dcos_w1} shows 
the $Q^2$-integrated differential cross sections
as a function of $|\cos \theta^*|$ for the four $W$ bins indicated in each panel.
The values of the S, D$_0$, D$_1$, and D$_2$ waves obtained in the nominal fit 
(at $\langle Q^2 \rangle = 6.5~\GeV^2$)
are shown for comparison.
It seems that the S wave is dominant in the energy region of $W$ near $1.1~\GeV$.
The amplitudes  $D_0$, $D_1$, and $D_2$ appear to be non-zero in the 
energy region of $W$ near $1.5~\GeV$; \textit{i.e.,} close to the mass of the 
$f_2'(1525)$ meson.

\begin{figure}
 \centering
  {\epsfig{file=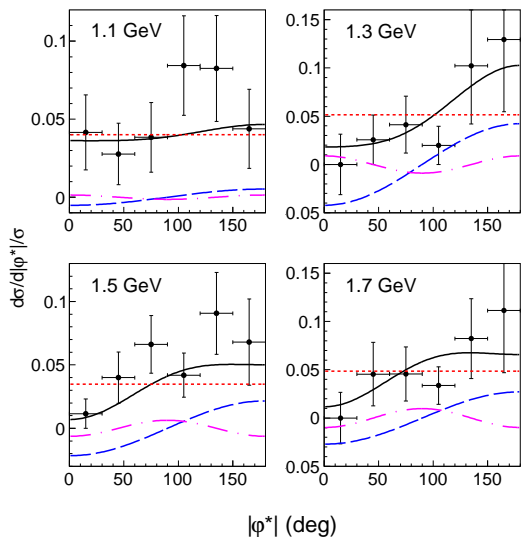,width=77mm}}
 \caption{
$|\varphi^*|$ dependence of the normalized differential cross sections and 
the fitted results in the four $W$ bins indicated in each panel.
The lines shown result from the nominal fit
(at $\langle Q^2 \rangle = 6.5~\GeV^2$).
Black solid line: total,
red dotted: $t_0$; blue dashed: $t_1 \cos |\varphi^*|$;
and magenta dash-dotted: $t_2 \cos 2 |\varphi^*|$.
}
 \label{fig:dphi_w1}
\end{figure}

Figure \ref{fig:dphi_w1} shows 
the $Q^2$-integrated differential cross sections
as a function of $|\varphi^*|$ for the four $W$ bins indicated in each panel.
The $t_0$, $t_1 \cos |\varphi^*|$, and $t_2 \cos 2 |\varphi^*|$ functions
obtained in the nominal fit 
(at $\langle Q^2 \rangle = 6.5~\GeV^2$)
are shown in the figure as well.

\begin{figure*}
\centering
\includegraphics[width=17cm]{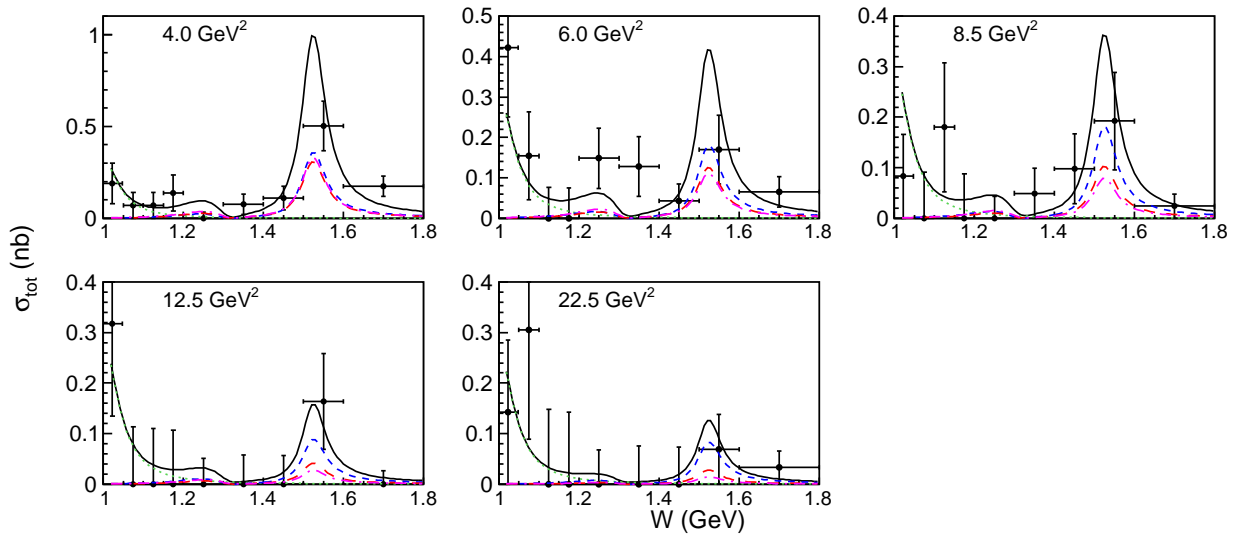}
\centering
 \caption{
Total cross sections (integrated over angle)
for $\gamma^* \gamma \to \ks \ks$ in five $Q^2$ bins 
as indicated in each panel, together with the
fit results described in \ref{sub:tff}.
Black solid line: total; 
green dotted: $|S|^2$; blue dashed: $|D_0|^2$;
red long-dashed: $|D_1|^2$; and magenta dash-dotted: $|D_2|^2$.
}
\label{fig:tot2}
\end{figure*}

The total cross sections (integrated over angle) for 
$\gamma^* \gamma \to \ks \ks$ are presented in Fig.~\ref{fig:tot2} in the five $Q^2$ bins
(in GeV$^2$) shown in each panel.
The results from the nominal fit are also shown.

\begin{figure}
 \centering
  {\epsfig{file=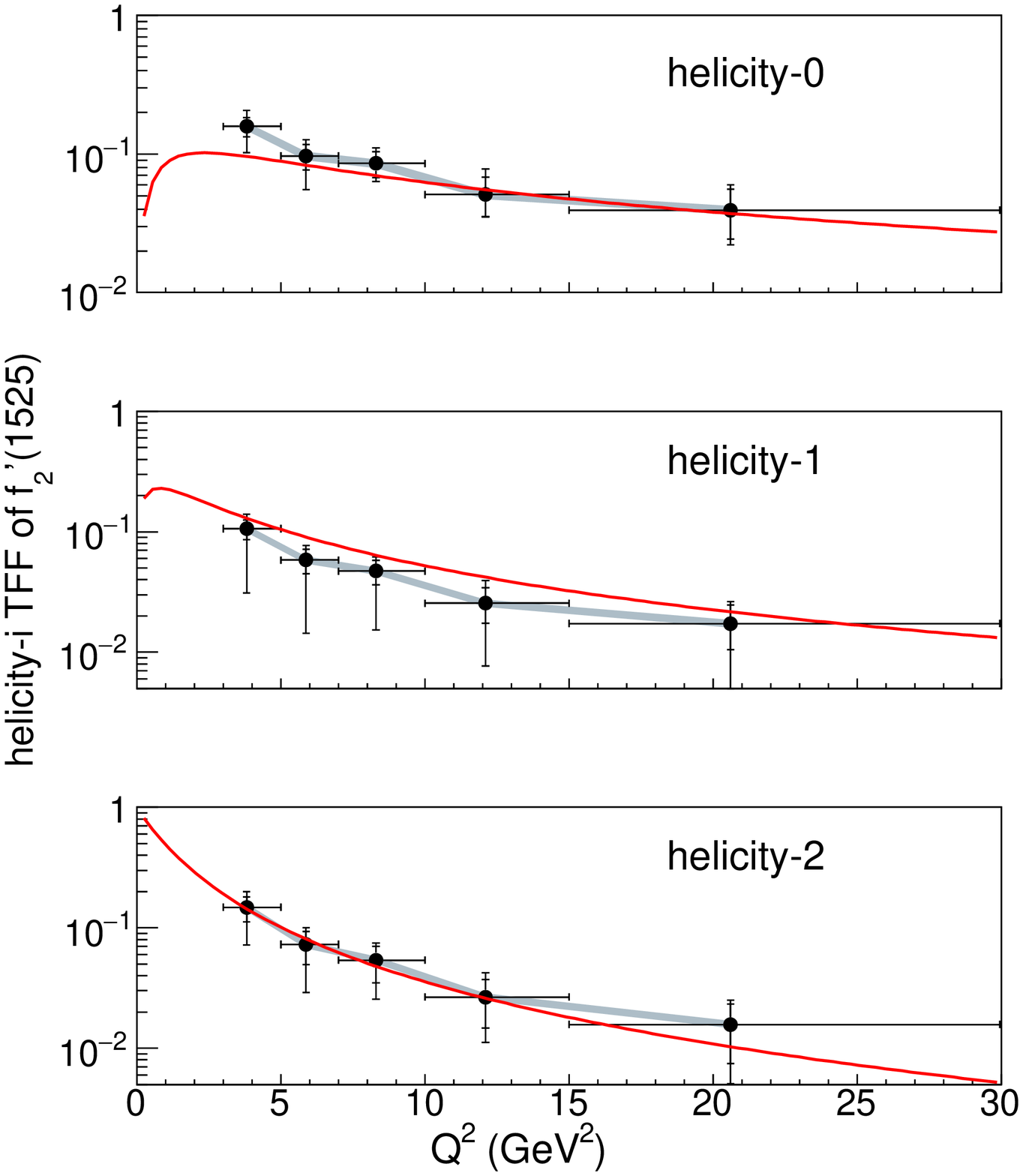,width=80mm}}
 \caption{The obtained helicity-0, -1, and -2 TFF of the $f_2'(1525)$ meson
as a function of $Q^2$, assuming Eq.~(\ref{eqn:r012}).
Short (long) vertical bars indicate statistical (combined statistical and systematic) errors.
The shaded band corresponds to the overall uncertainty arising from the known errors on
$\Gamma_{\gamma\gamma}$.
The solid line shows the predicted $Q^2$ dependence in 
SBG~\cite{schuler}.}
 \label{fig:fi2q2}
\end{figure}

The obtained $Q^2$ dependences of the helicity-0, -1, and -2 TFF, 
$\sqrt{r_{ifp}} F_{f2p}$ ($i= 0$, 1, 2), 
for the $f_2'(1525)$ meson obtained from
the nominal fit are shown in Table~\ref{tab:tff_f2p} and Fig.~\ref{fig:fi2q2}.
Also shown is the $Q^2$ dependence predicted by SBG~\cite{schuler}.
Note that we have assumed Eq.~(\ref{eqn:r012}) in the fit, without which 
fits often fail due to the limited statistics.
With this caveat, the measured helicity-0 and -2 TFFs of 
the $f_2'(1525)$ meson
agree well with SBG~\cite{schuler} and the helicity-1 TFF is not 
inconsistent with prediction.

\begin{center}
\begin{table*}
\caption{Fitted parameters of cross sections and the number of solutions
obtained under the conditions noted below.
In each category, only solutions assuming $k_0\ne0 \ \bigcap\ k_1\ne0$ 
are shown.
Only the single solution that gives the minimum $\chi^{2}$ in category 3 
is shown, while two viable solutions in categories 1 and 2 are shown.}
\label{tab:comfit}

\begin{tabular}{l|cc|cc|c} \hline \hline
Parameter
 & \multicolumn{2}{c|}{Category 1}
 & \multicolumn{2}{c|}{Category 2}
 & \multicolumn{1}{c}{Category 3} \\ 
\hline
Conditions & \multicolumn{2}{c|}{$A_{BW}\ne 0 \ \bigcap\ B_S=0 $} & \multicolumn{2}{c|}{$A_{BW}=0 \ \bigcap\ B_S\ne 0 $}& $A_{BW}=B_{S}=0$ \\

\hline
Number of solutions &\multicolumn{2}{c|}{2}  & \multicolumn{2}{c|}{2}  & 3  \\
 & Solution 1a &Solution 1b  & Solution 2a & Solution 2b  &   \\
\hline
$\chi^2_{\rm P}/ndf$ &152.4/150 &159.8/150 & 154.9/151 & 156.1/151  & 293.9/155    \\ \hline

$k_0$~(GeV$^{-2}$) &$0.30^{+0.31}_{-0.14}$ &$0.31^{+0.34}_{-0.15}$ & $0.31^{+0.34}_{-0.15}$ & $0.29^{+0.31}_{-0.14}$    & $0.33^{+0.31}_{-0.14}$        \\
$k_1$~(GeV$^{-1}$) &$0.27^{+0.30}_{-0.14}$ &$0.27^{+0.44}_{-0.15}$ & $0.29^{+0.33}_{-0.15}$ & $0.24^{+0.29}_{-0.13}$   & $0.23^{+0.25}_{-0.12}$    \\
\hline
$F_{f2p}(0.0);(\times 10^{-2})$ & \multicolumn{5}{c}{$100 \pm 7$  } \\

$F_{f2p}(4.0);(\times 10^{-2})$ &$24.1^{+2.6}_{-2.5}$ &  $24.4^{+2.7}_{-2.6}$ & $24.3^{+2.6}_{-2.5}$  & $24.4^{+2.6}_{-2.5}$  &  $27.1^{+2.7}_{-2.6}$ \\
$F_{f2p}(6.0);(\times 10^{-2})$ &$13.4^{+2.6}_{-2.5}$ &  $13.9^{+2.5}_{-2.4}$ & $14.3^{+2.5}_{-2.3}$  & $14.4^{+2.5}_{-2.3}$  &  $15.5^{+2.5}_{-2.4}$  \\
$F_{f2p}(8.5);(\times 10^{-2})$ &$11.2^{+2.3}_{-2.2}$ &  $11.3^{+2.3}_{-2.2}$ & $11.5^{+2.3}_{-2.2}$  & $11.6^{+2.3}_{-2.1}$  &  $12.4^{+2.3}_{-2.2}$  \\
$F_{f2p}(12.5);(\times 10^{-2})$ &$6.3^{+2.1}_{-1.9}$ &  $6.3^{+2.1}_{-1.9}$  & $6.3^{+2.1}_{-1.9}$   & $6.3^{+2.1}_{-1.9}$  &  $7.0^{+2.1}_{-1.9}$   \\
$F_{f2p}(22.5);(\times 10^{-2})$ &$4.6^{+1.9}_{-1.7}$ &  $4.6^{+1.9}_{-1.7}$  & $4.6^{+1.9}_{-1.7}$   & $4.7^{+1.9}_{-1.7}$  &  $5.1^{+2.0}_{-1.8}$  \\
\hline

$\phi_{fpD1}~(^\circ);$ &$33^{+28}_{-81}$ & $177^{+27}_{-27}$   & $112^{+23}_{-35}$  & $108^{+24}_{-37}$  & $47^{+24}_{-33}$   \\
$\phi_{fpD2}~(^\circ);$ &$199^{+34}_{-75}$ & $218^{+27}_{-29}$  & $209^{+30}_{-35}$  & $213^{+28}_{-33}$  & $218^{+23}_{-27}$    \\
$\phi_{faD1}~(^\circ);$ &$137^{+27}_{-34}$ &$328^{+34}_{-39}$   & $18^{+28}_{-30}$   & $340^{+33}_{-33}$  & $234^{+22}_{-24}$   \\
$\phi_{faD2}~(^\circ);$ &$166^{+30}_{-32}$ & $180^{+29}_{-29}$  & $162^{+29}_{-32}$  & $182^{+27}_{-28}$  & 0 (fixed)  \\
\hline
$f_S$~($\sqrt{\rm nb}~$GeV$^2$);$(\times 10^{-2})$  &  $1.3^{+1.1}_{-0.6}$  &  $0.9^{+0.8}_{-0.4} $  & \multicolumn{2}{c|}{ 0 (fixed) }   & 0 (fixed)     \\
$g_S$~(GeV) &  $0.10^{+0.05}_{-0.04}$ &  $0.06^{+0.05}_{-0.05} $  &  \multicolumn{2}{c|}{ 0 (fixed)}   & 0 (fixed)     \\
$p_S$ &  $0.06^{+0.25}_{-0.24}$ &  $0.01^{+0.26}_{-0.25} $  &  \multicolumn{2}{c|}{ 0 (fixed)}   & 0 (fixed)     \\
$\phi_{BW}~(^\circ)$; & $297^{+21}_{-21}$ & $150^{+35}_{-24}$ &  \multicolumn{2}{c|}{ 0 (fixed)}  & 0 (fixed)  \\

\hline
$a_S~(\sqrt{\rm nb});(\times 10^{-3})$ & \multicolumn{2}{c|}{0 (fixed) }  & $4.3^{+12.5}_{-5.9}$ & $2.2^{+5.7}_{-3.0}$ &  0 (fixed)     \\
$b_S$ & \multicolumn{2}{c|}{0 (fixed) } & $19.6^{+4.6}_{-4.1}$   &  $21.9^{+6.0}_{-4.0}$     &  0 (fixed)  \\
$c_S$ & \multicolumn{2}{c|}{0 (fixed)}  & $0.00^{+0.23}_{-0.06}$ &  $0.00^{+0.21}_{-0.05}$   & 0 (fixed)      \\
$\phi_{BS}~(^\circ)$; & \multicolumn{2}{c|}{  0 (fixed)} &  $99^{+19}_{-21}$ & $311^{+20}_{-18}$ &  0 (fixed) \\

\hline \hline
\end{tabular}
\end{table*}
\end{center}

\begin{center}
\setlength{\myheight}{4mm}
\begin{table*}
\caption{
Transition form factors of the $f_2'(1525)$ meson ($\times 10^{-2}$) 
for each helicity and combined.
The first and second uncertainties are statistical and systematic, respectively.
The normalization of the TFF is such that $F_{f2p}(0)=1$.
There is an additional overall systematic uncertainty of $\pm 7\%$ due to
the error in the tabulated two-photon decay width $\Gamma_{\gamma\gamma}$ of the 
$f_2'(1525)$ state.
}
\label{tab:tff_f2p}
\begin{tabular}{c|cccc} \hline \hline
\rule{0cm}{\myheight} $\overline{Q}^2 (\GeV^2) $ & helicity-0 & helicity-1 & helicity-2 & Total \\ 
\hline \hline

3.51 &$15.8^{+2.4}_{-2.5}$$^{+4.1}_{-5.1}$  &$10.6^{+1.9}_{-2.0}$$^{+2.8}_{-7.3}$  &$14.8^{+3.3}_{-3.6}$$^{+4.1}_{-6.6}$ &$24.1^{+2.6}_{-2.5}$$^{+6.0}_{-8.2}$\\ 
5.87 &$9.7^{+2.0}_{-2.0}$$^{+2.2}_{-3.6}$  &$5.8^{+1.3}_{-1.4}$$^{+1.3}_{-4.2}$  &$7.3^{+2.1}_{-2.3}$$^{+1.8}_{-3.7}$ &$13.4^{+2.6}_{-2.5}$$^{+3.0}_{-5.2}$\\ 
8.30 &$8.6^{+1.8}_{-1.8}$$^{+1.7}_{-1.2}$  &$4.7^{+1.1}_{-1.1}$$^{+1.0}_{-3.0}$  &$5.4^{+1.6}_{-1.9}$$^{+1.3}_{-2.1}$ &$11.2^{+2.3}_{-2.2}$$^{+2.2}_{-1.8}$\\ 
12.1 &$5.1^{+1.7}_{-1.6}$$^{+2.1}_{-0.0}$  &$2.6^{+0.9}_{-0.8}$$^{+1.1}_{-1.6}$  &$2.7^{+1.1}_{-1.2}$$^{+1.2}_{-1.0}$ &$6.3^{+2.1}_{-1.9}$$^{+2.6}_{-0.3}$\\
20.6 &$3.9^{+1.7}_{-1.5}$$^{+1.2}_{-0.9}$  &$1.7^{+0.7}_{-0.7}$$^{+0.5}_{-1.1}$  &$1.6^{+0.8}_{-0.8}$$^{+0.5}_{-0.7}$ &$4.6^{+1.9}_{-1.7}$$^{+1.4}_{-1.0}$\\

\hline \hline
\end{tabular}
\end{table*}
\end{center}

\subsection{Estimation of systematic uncertainties of the TFF}
\label{sub:systff}
In this subsection, we estimate systematic uncertainties for the TFF of the
$f_2'(1525)$ meson.
These arise primarily from the overall
$\pm 7$\% normalization uncertainty on $\Gamma_{\gamma \gamma}$
that affects all $Q^2$ bins uniformly and the individual uncertainties that vary in each $Q^2$ bin.
The individual systematic uncertainties evaluated below
are converted to uncertainties in the helicity-0, -1, and -2 
components of the TFF
of the $f_2'(1525)$ meson as summarized in Table~\ref{tab:tff_f2p} and shown in 
Fig.~\ref{fig:fi2q2}. 
All uncertainties are summed quadratically in each $Q^2$ bin to obtain the total systematic error in that bin.

Individual uncertainties are estimated for the TFF as follows.
The uncertainties of the normalization factor in the differential 
cross sections 
are estimated by shifting the value corresponding to 1$\sigma$ of the fit.
The systematic uncertainties of the measured 
total cross sections 
are taken into account by refitting the cross sections 
with the error shifted.
The properties such as the mass, the width, and the branching fraction to 
$K \bar{K}$ of the $f_2(1270)$, the $a_2(1320)$, and the $f_2'(1525)$ mesons
are shifted by the uncertainties given in the PDG~\cite{pdg2016}.
The $m_0^2$ in $A_{BW}$ is changed to $(1.0 \pm 0.5)~\GeV^2$.
For $B_{Di}$, they are turned on individually and their effects 
are taken as uncertainties.

Systematic uncertainties due to various possible distortions in the
distributions of $W$, $Q^2$, $|\cos \theta^*|$, and $|\varphi^*|$ studied 
below are evaluated parametrically.
The effect of a shift of $\pm 10$\% in the total and the differential cross sections
over the full range of $W$ 
is estimated by multiplying the cross sections by 
[$1 \pm 0.25\times(W-1.4~\GeV)$]. 
The effect of a shift of $\pm 5$\% in the total cross sections
over the full range of $Q^2$ is evaluated
by multiplying by [$1 \pm 0.006\times(Q^2-12.2~\GeV^2)$].
Additional uncertainties considered are those arising from changing the range of $W$,
from $1.0-1.8$ to $1.0-2.0$ or $1.0-1.6$~GeV.
The effect of a shift of $\pm 10$\% in the differential cross sections
as a function of $\cos \theta^*$ is evaluated by multiplying by 
[$1 \pm 0.2\times(|\cos \theta^*|-0.5)$].
The effect of a shift of $\pm 10$\% in the differential cross sections 
as a function of $|\varphi^*|$ 
is evaluated by multiplying by 
[$1 \pm 0.0011\times(|\varphi^*|-90^\circ)$].
The uncertainty in the convex or concave shape of $\cos \theta^*$ is
evaluated by multiplying by [$1.1- 0.8\times(|\cos \theta^*|-0.5)^2$],
or [$0.9 + 0.8 \times(|\cos \theta^*|-0.5)^2$], respectively.
Similarly, the uncertainty in the convex or concave shape of $|\varphi^*|$ is evaluated by multiplying by
[$1.1 - 2.5\times10^{-5}\times(|\varphi^*|-90^\circ)^2$] or 
[$0.9 + 2.5\times10^{-5}\times(|\varphi^*|-90^\circ)^2$], respectively.

\subsection{$Q^2$ dependence of cross sections near the \boldmath{$\ks\ks$} threshold}
In the $\gamma\gamma \rightarrow \ks\ks$ reaction, 
a peak structure near $\ks\ks$ threshold is expected, based on a comprehensive 
amplitude analysis using the data of $\gamma\gamma \rightarrow \pi\pi$ and 
$K \bar{K}$~\cite{Pennington}.
In Refs.~\cite{achasov1} and \cite{achasov2}, it is predicted that 
this peak structure persists even if the $f_0(980)$ and the $a_0(980)$ mesons 
interfere destructively.
Experimentally, there have been no measurements to date of the
two-photon cross section 
in the energy region of $W$ below $1.05~\GeV$.

\begin{figure}
 \centering
  {\epsfig{file=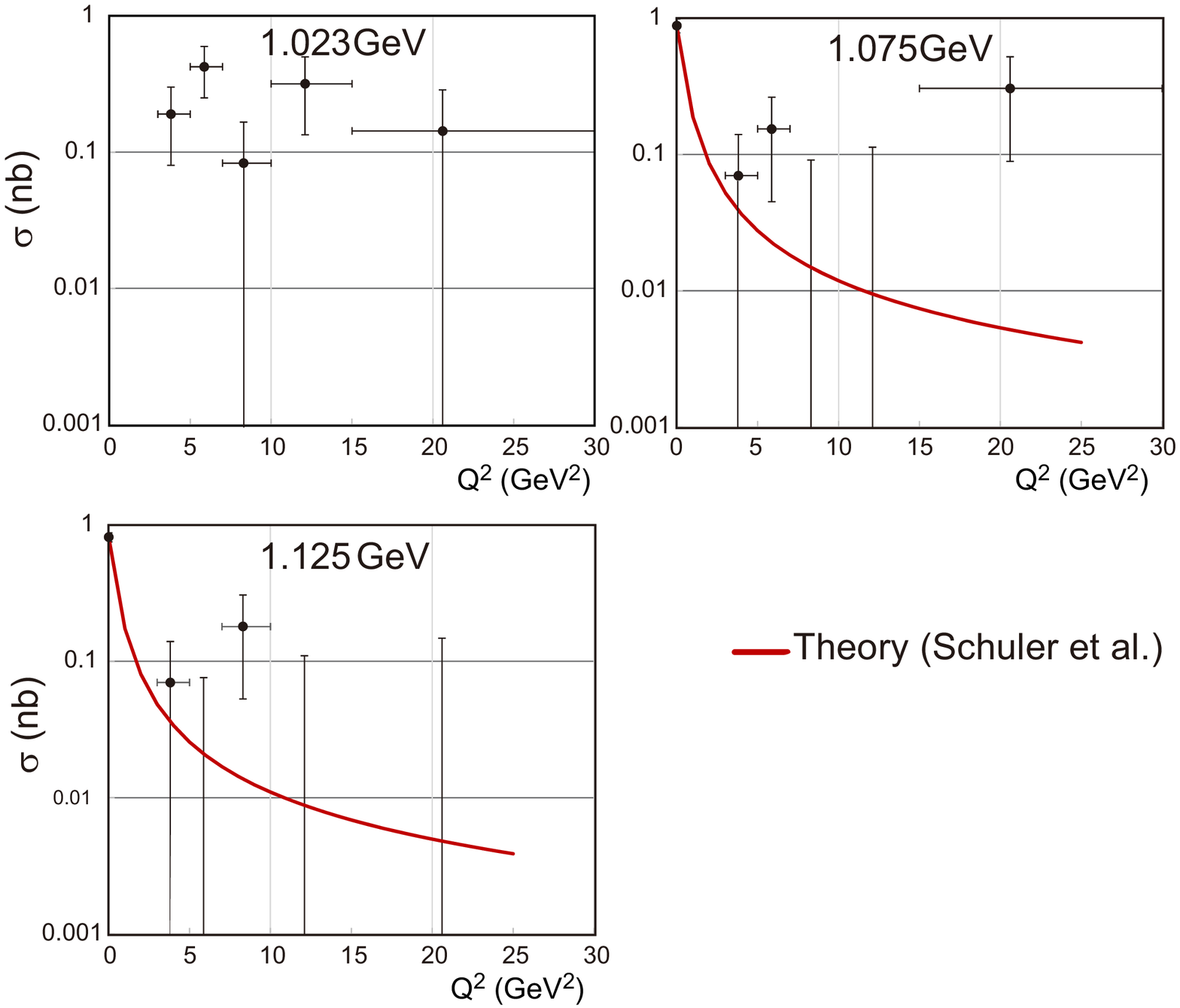,width=77mm}}
 \caption{$Q^2$ dependence of the cross section in the 
three $W$ regions near the $\ks\ks$ mass threshold,
with central values as indicated in the subpanels.
Only statistical errors are shown.
Solid curves show the predicted $Q^2$ dependence in
SBG~\cite{schuler}}.
\label{fig:thq2}
\end{figure}

The nominal fit shows that S wave can be expressed 
by a Breit-Wigner function with a mass of 0.995 GeV/$c^2$. 
Motivated by this, we have plotted the $Q^2$ dependence of the total cross
sections in the energy bins at 1.023 GeV, 1.075 GeV, and 1.125 GeV as shown in 
Fig.~\ref{fig:thq2}.
We also show the $Q^2$ dependence for a $J^P=0^+$ state predicted with 
$M=0.98~\GeV/c^2$ in SBG~\cite{schuler}
normalized by the points at $Q^2 = 0$, which are translated from
the data of the no-tag measurement of this process~\cite{ksks} 
assuming an isotropic angular dependence. 
These data are available at the two higher-$W$ regions. 
The measured cross sections are slightly larger than the predicted values,
though not inconsistent with them given the large statistical errors.
The cross sections increase as $W$ approaches the mass threshold,
which may signify the threshold enhancement suggested in Ref.~\cite{Pennington}. 

\section{Summary and Conclusion}
\label{sec:summary}
We have measured the cross section of $\ks$-pair production in single-tag 
two-photon collisions, $\gamma^* \gamma \to \ks \ks$ up to $Q^2 = 30~\GeV^2$
based on a data sample of 759~fb$^{-1}$ collected
with the Belle detector at the KEKB asymmetric-energy 
$e^+ e^-$ collider.
The data covers the kinematic range $1.0~\GeV < W < 2.6~\GeV$ and the angular 
range of $|\cos \theta^*| < 1.0$ and $0 \le |\varphi^*| \le 180^\circ$
in the $\gamma^* \gamma$ c.m. system. 

For the first time, we find the
$f_2'(1525)$, $\chi_{c0}(1P)$, and $\chi_{c2}(1P)$ mesons in high-$Q^2$ 
$\gamma^* \gamma$ scattering.
These resonances are most visible
in the corresponding no-tag mode~\cite{ksks}.

We have estimated the 
$\chi_{c0}$ and $\chi_{c2}$ partial decay widths $\Gamma_{\gamma*\gamma}$ 
as a function of $Q^2$.
The $Q^2$ dependences of $\Gamma_{\gamma*\gamma}$ are normalized to 
$\Gamma_{\gamma\gamma}$ at $Q^2 = 0$ and compared with 
SBG~\cite{schuler},
as shown in Fig.~\ref{fig:charmonium_ggg}.
They are in agreement, albeit with very limited statistics.

A partial-wave analysis has also been conducted for the $\gamma^* \gamma \to \ks \ks$ event sample.
The helicity-0, -1, and -2 transition form factors (TFFs) of 
the $f_2'(1525)$ meson 
are measured for the first time for $Q^2$ up to $30~\GeV^2$ 
and are compared with theoretical predictions.
The measured helicity-0 and -2 TFFs of the $f_2'(1525)$ meson agree well with 
SBG~\cite{schuler}, and
the helicity-1 TFF is not inconsistent with prediction.

We have also compared the total cross section near the $\ks\ks$ mass threshold
as a function of $Q^2$ with the prediction for a $J^P=0^+$ state with 
$M=0.98~\GeV/c^2$ in SBG~\cite{schuler}, although our limited statistics currently
preclude a conclusive interpretation.

\section*{Acknowledgments}
% Please paste this acknowledgement into your latex file.
%
% 2017.08.31 updated Korea
% 2017.05.25 updated Spain
% 2017.03.17 updated Korea, Czechia
% 2017.01.26 updated SINET5 (short only) 
% 2016.10.15 added China new No., removed MEXT+JSPS (Iijima-san),
% 	     Austria, updated SINET5 and Korean (Long only) 
% 2015.12.17 added China new No., Germany EXC, Taiwan, Spain
% 2015.08.25 updated Korea
% 2015.08.11 removed Australian Department of Industry, Innovation, 
%                    Science and Research (long); DIISR (short)
% 2015.04.27 add CCEPP (China)
% 2014.11.14 spelled out Korea & Spain FA (APS request, Long only)
% 2014.09.30 updated China grant #s (Long only)
% 2014.09.01 delete WCU (both). added three contract No.'s and 
%            delete one contract No.'s in long version
% 2014.07.28 added grant No. P 26794-N20 for Austria
% 2014.03.31 corrected "WCU program of the Ministry <of> Education"
% 2014.02.25 updated Czech contract no. (Long only)
% updated  2014.01.29 added Center for Korean J-PARC Users (Long only)
% updated  2013.11.20 included mods done by APS editor (Long only)
% updated  2013.09.19 new Korean part
% updated  2013.07.31 added two contract No.'s for China (Long only)
%
%***** Acknowledgments *****

%----------- Long version, for most papers ----------- 
We thank the KEKB group for the excellent operation of the
accelerator; the KEK cryogenics group for the efficient
operation of the solenoid; and the KEK computer group,
the National Institute of Informatics, and the 
PNNL/EMSL computing group for valuable computing
and SINET5 network support.  We acknowledge support from
the Ministry of Education, Culture, Sports, Science, and
Technology (MEXT) of Japan, the Japan Society for the 
Promotion of Science (JSPS), and the Tau-Lepton Physics 
Research Center of Nagoya University; 
the Australian Research Council;
Austrian Science Fund under Grant No.~P 26794-N20;
the National Natural Science Foundation of China under Contracts 
No.~10575109, No.~10775142, No.~10875115, No.~11175187, No.~11475187, 
No.~11521505 and No.~11575017;
the Chinese Academy of Science Center for Excellence in Particle Physics; 
the Ministry of Education, Youth and Sports of the Czech
Republic under Contract No.~LTT17020;
the Carl Zeiss Foundation, the Deutsche Forschungsgemeinschaft, the
Excellence Cluster Universe, and the VolkswagenStiftung;
the Department of Science and Technology of India; 
the Istituto Nazionale di Fisica Nucleare of Italy; 
National Research Foundation (NRF) of Korea Grants No.~2014R1A2A2A01005286, 
No.~2015R1A2A2A01003280,No.~2015H1A2A1033649, No.~2016R1D1A1B01010135, 
No.~2016K1A3A7A09005603, No.~2016R1D1A1B02012900; Radiation Science Research Institute, 
Foreign Large-size Research Facility Application Supporting project 
and the Global Science Experimental Data Hub Center of the Korea Institute of Science and Technology Information; 
the Polish Ministry of Science and Higher Education and 
the National Science Center;
the Ministry of Education and Science of the Russian Federation 
under contracts No. 3.3008.2017/PP and other contracts 
and the Russian Foundation for Basic Research;
the Slovenian Research Agency;
Ikerbasque, Basque Foundation for Science and
MINECO (Juan de la Cierva), Spain;
the Swiss National Science Foundation; 
the Ministry of Education and the Ministry of Science and Technology of Taiwan;
and the U.S.\ Department of Energy and the National Science Foundation.

%\newpage


\begin{thebibliography}{99}
\bibitem{pw} We denote individual partial
waves by upright letters and parameterized waves by italics.
\end{thebibliography}
\end{document}